\documentclass[seceq]{ptptex}
\usepackage{wrapft,graphicx,amsmath,amssymb}
\notypesetlogo
\title{
  Deformations and clustering correlations in $p$- and $sd$-shell nuclei using the Gogny and Skyrme interactions
} 
\author{
  Yasutaka \textsc{Taniguchi}, Yoshiko \textsc{Kanada-En'yo}$^*$ and Masaaki \textsc{Kimura}$^{**}$
}
\inst{
  Research Center for Nuclear Physics, Osaka University, Ibaraki 567-0047, Japan\\
  $^*$Yukawa Institute for Theoretical Physics, Kyoto University, Kyoto 606-8502, Japan\\
  $^{**}$Creative Research Initiative ``Sousei'', Hokkaido University, Sapporo 001-0021, Japan
}
\abst{
  Structures of $p$- and $sd$-shell nuclei are studied
  with the deformed-basis antisymmetrized molecular dynamics method 
  using the Gogny D1S and Skyrme SLy7 forces as effective interactions.
  By the energy variation with a constraint, energy curves 
  as functions of quadrupole deformation parameter $\beta$ are obtained.
  The energy curves for $sd$-shell nuclei
  show structure change as a function of $\beta$, and suggest shape coexistence. 
  Nuclear structures in the deformed region are discussed 
  focusing on deformations and clustering.
  It is found that the deformations often involve cluster structures.
  Effective-interaction dependence is also discussed comparing the results 
  obtained with the two effective interactions.
  Although the two forces give similar results when cluster structures do not 
  develop clearly, they give different energies in the
  largely deformed region when an $\alpha$ cluster develops. 
}
  
\begin{document}
  \newcommand{\bra}[1]{\langle #1 |}
  \newcommand{\ket}[1]{| #1 \rangle}
  \newcommand{\vect}[1]{\mathbf #1}
  \newcommand{\ovlp}[2]{\langle #1 | #2 \rangle}
  \newcommand{\expec}[1]{\mbox{$\left\langle #1 \right\rangle$}}
  \newcommand{\norm}[1]{\mbox{$\bra{ #1 }\left.\! #1 \right>$}}
  \newcommand{\inpro}[2]{\mbox{$\bra{ #1 }\left.\! #2 \right>$}}
  \newcommand{\Phirm}{{\rm \Phi}}
  \newcommand{\EPJA}[3]{\JL{Eur. Phys. J.,A#1,#2,#3}}
  \maketitle

  \section{Introduction}
  \label{sec:introduction}
 It is well known that a variety of structures appears in excited states
as well as ground states.
Shape coexistence is one of the typical examples 
that a variety of structures coexists in a nucleus.
Various deformed structures have been observed in low-excited states
of nuclei in medium- and heavy-mass regions.
Recently, superdeformed rotational band have been 
observed in $^{36}$Ar\cite{36ArSD} and $^{40}$Ca\cite{40CaSD}. 
Motivating the observation, microscopic studies for deformed structures in $sd$-shell region have been progressed nowadays. 
For instance, Inakura {\it et al.} have studied deformations
in $sd$- and $pf$-shell region systematically with 
the Hartree-Fock method using four parameter sets of 
the Skyrme forces, and suggested superdeformations and 
hyperdeformations in these mass region\cite{ina02}. 
Moreover, the Hartree-Fock-Bogoliubov (or -BCS) + generator coordinate method with respect to quadrupole deformation have been carried out in $sd$-shell region\cite{rod00,rod02,ben03a,rod03,rod04,egi04}. 
Thus, nuclear deformation plays an important role also 
in the $sd$-shell nuclei, and searching for 
exotic deformations such as superdeformation in this region
is one of hot subjects.

Another important aspect in the light-mass region is cluster 
structure. It has been known that cluster structures often appear
in excited states and also in the ground states of some nuclei 
in $p$-shell and light $sd$-shell region\cite{hor72,fuj80}.
Recently, even in heavy $sd$-shell nuclei such as $^{32}$S\cite{kim04b} and $^{40}$Ca\cite{kan05,tan07}, 
the cluster aspect has been found to be important in 
low-lying deformed states in full microscopic studies. 
In other words, largely deformed states often involve clustering at least
in $Z=N$ nuclei in this mass region. We should point out here that 
the clustering in the low-lying deformed states is not always well marked with an enhanced intercluster distance, 
but it has spatial correlation of nucleons forming cluster cores.
In this sense, we call both the 
formation of cluster cores and the spatially developed clustering 
``clustering correlations''.

As mentioned above, both the deformations and the cluster correlations
play important roles in the light-mass region. 
In the theoretical studies, it is important to 
investigate clustering correlations in 
deformed states systematically. 
However, such studies have not been progressed enough.

Our aim is to study structure of deformed states and roles of clustering there. 
For this aim, we study deformations and clustering correlations in $p$- and $sd$-shell $N=Z=\mbox{even}$ nuclei and a neutron-rich nucleus $^{24}$O systematically, and analyze structure changes as functions of 
the quadrupole deformation.
We use the framework of the deformed-basis AMD\cite{kim04a}, which is suitable
to study deformations and cluster correlations. 
The deformed-basis AMD wave function is constructed by a Slater determinant of triaxially deformed Gauss' wave packets. In this approach, 
mean-field deformations are described by deformations of wave packets while
clustering are expressed by localizing of centers of the wave packets.
We note that the deformed-basis AMD is powerful framework to study deformations and clustering in a unified manner as already proved in the previous
studies of $sd$-shell nuclei 
\cite{kim04a,kim04b,kan05,tan04,tan07,kim07}. 
We perform energy variation imposing constraint on 
quadrupole deformation parameter $\beta$ and analyze the energy curves
and structure changes as functions of $\beta$, focusing on
cluster correlation in the deformed states.
We use two effective interactions, the Gogny D1S (D1S)\cite{D1S} and Skyrme SLy7 (SLy7)\cite{cha98} forces.
The former is a finite-range interaction and the latter is a zero-range
interaction. Comparing the results obtained by the two forces, 
the interaction dependences are also discussed.

  The paper is organized as follows: Section~\ref{sec:framework} gives a brief outline of the deformed-basis AMD method and properties of the D1S and SLy7 forces. 
  In Sec.~\ref{sec:deformations}, the calculated results are shown, and features of deformations in $p$- and $sd$-shell nuclei are described. 
 The interaction dependences are also discussed. 
  In Sec.~\ref{sec:clustering}, we analyze the deformed structures of the $p$- and $sd$-shell nuclei focusing on cluster aspects.
  Section~\ref{sec:conclusion} is devoted to a conclusion. 
  
  \section{Framework}
  \label{sec:framework}
  \subsection{Wave function and energy variation with constraint in the deformed-basis AMD}
  We use the theoretical framework of the deformed-basis AMD method whose wave function is a Slater determinant of triaxially deformed Gauss' wave packets, 
  \begin{subequations}
    \label{AMD}
    \begin{eqnarray}
      \ket{\Phirm_{\rm int}}& = &\hat{\cal A} \ket{\varphi_1,\  \varphi_2,\cdots,\varphi_A}, \\
      \ket{\varphi_i}& = &\ket{\phi_i,\  \chi_i,\   \tau_i}, \\
      \ovlp{\vect{r}}{\phi_i}& = &\prod_{\sigma = x, y, z} \left( \frac{2\nu_\sigma}{\pi} \right)^{\frac{1}{4}} \exp \left[ - \nu_\sigma \left( r_\sigma - \frac{Z_{i\sigma}}{\sqrt{\nu_\sigma}} \right)^2 \right], \nonumber\\
      \\
      \ket{\chi_i}& = &\alpha_i \ket{\uparrow} + \beta_i \ket{\downarrow},\\
      \ket{\tau_i}&  = &\ket{p}\  {\rm or}\  \ket{n}. 
    \end{eqnarray}
  \end{subequations}
  Here, the complex parameters $\vect{Z}_i$, which represent the centroids
  of the Gauss' wave packets in phase space, take independent values for each single-particle wave function.  The width parameters $\nu_x$, $\nu_y$ and
  $\nu_z$ are real parameters and take independent values for each of the
  $x$-, $y$- and $z$-directions, but are common for all nucleons.  The
  spin part $\ket{\chi_i}$ is parametrized by $\alpha_i$ and $\beta_i$ and the
  isospin part $\ket{\tau_i}$ is fixed to $\ket{p}$ (proton) or $\ket{n}$ (neutron).  The
  quantities $\{\vect{Z}_i,\alpha_i,\beta_i,\nu_x,\nu_y,\nu_z\}$ are
  variational parameters and are optimized by energy variation as
  explained following. 
  
  A trial wave function in the energy variation with constraints is
  a parity-projected wave function,  
  \begin{equation}
    \ket{\Phirm^\pi} = \frac{1+\pi \hat{P}_r}{2} \ket{\Phirm_{\rm int}}, 
  \end{equation}
  where $\pi$ is a parity, and $\hat{P}_r$ is the parity operator. 

  The Hamiltonian is,
  \begin{equation}
    \hat{H} = \hat{K} + \hat{V}_{\rm N} + \hat{V}_{\rm C} - \hat{K}_{\rm G}, 
  \end{equation}
  where $\hat{K}$ and $\hat{K}_{\rm G}$ are the kinetic energy and the
  energy of the center of mass motion, respectively, and $\hat{V}_{\rm N}$
  is the effective nucleon-nucleon interaction. 
  The Gogny D1S and Skyrme SLy7 forces are used in the present work. 
  The Coulomb force
  $\hat{V}_{\rm C}$ is approximated by a sum of seven Gauss' functions.  
  
  We perform energy variation and optimize the variational parameters
  to find a state that minimizes the energy of the system $E^\pi$,  
  \begin{equation}
    E^\pi = \frac{\bra{\Phirm^\pi}\hat{H}\ket{\Phirm^\pi}}{\ovlp{\Phirm^\pi}{\Phirm^\pi}} + V_{\rm cnst}. 
  \end{equation}
  Here, we add the constraint potential $V_{\rm cnst}$ to the expectation
  value of Hamiltonian $\hat{H}$ in order to obtain local minimum energy
  states under a constraint.  
  In this study, we employ  constraint on the  quadrupole
  deformation parameter $\beta$\cite{dot97}, 
  \begin{equation}
    V_{\rm cnst} = v_{\rm cnst} ( \beta - \tilde{\beta} )^2. 
  \end{equation}
  When sufficiently large value is chosen for $v_{\rm cnst}$, the resultant values of $\beta$ become $\tilde{\beta}$. 
  In present study, other quantities such as quadrupole deformation parameter $\gamma$ or octupole deformation are not constrained. 
  Therefore, those quantities are optimized for a given $\beta$ value by energy variation. 
  
  The energy variation with the AMD wave function is carried out using the
  frictional cooling method\cite{kan95}.  
  The time evolution equations for the complex variational parameters $\vect{Z}_i, \alpha_i$ and $\beta_i$ are  
  \begin{equation}
    \frac{dX_i}{dt} = - \mu_X \frac{\partial E^\pi}{\partial X_i^*},\  (i=1, 2, \cdots, A), 
  \end{equation}
  where $X_i$ is $\vect{Z}_i, \alpha_i$ or $\beta_i$, and those for the
  real parameters $\nu_x, \nu_y$ and $\nu_z$ are  
  \begin{equation}
    \frac{d\nu_\sigma}{dt} = - \mu_\nu \frac{\partial E^\pi}{\partial
      \nu_\sigma},\  (\sigma = x, y, z).  
  \end{equation}
  The quantities $\mu_X$ and $\mu_\nu$ are arbitrary positive real numbers. 
  The energy of the system decreases as time progresses, and after a
  sufficient number of time steps, a local minimum energy state is obtained under the given constraint.  

  \subsection{Effective interactions}
  \label{chap:interaction}

  We use the Gogny D1S and Skyrme SLy7 forces in the present study, which are widely used interactions in microscopic studies of nuclear structure. 
  The Skyrme force is zero-range and a density dependent force, and has been widely used to self-consistent nuclear structure calculations due to its numerical simplicity. 
  It is allocated to make density functionals. 
  The Gogny force consists of a finite-range part with Gaussian forms and a zero-range density dependent part. 
When the Gogny force is used in the Hartree-Fock calculations, single-particle wave functions are often expanded to harmonic oscillator basis practically. 
Although fewer studies using the Gogny force have been carried out than those using the Skyrme force because of more complicated form, it is confirmed that the Gogny force is a credible effective interaction by HFB and AMD\cite{sugawa_kimura} calculations. 

\begin{table}[tbp]
\begin{center}
\caption{Properties of infinite nuclear matter. }
\label{tab:matter}
\begin{tabular}{llccc}
\hline
Force & Parameter set & $\rho_{0, {\rm eq}}$ [fm$^{-3}$] & $K_\infty$ [MeV] & $m^*_0/m$ \\
\hline
Gogny & D1S & 0.160 & 209 & 0.67 \\
Skyrme & SLy7 & 0.158 & 230 & 0.69 \\
&SIII & 0.145 & 356 & 0.76 \\
\hline
\end{tabular}
\end{center}
\end{table}
Let us show the properties of infinite nuclear matter for the D1S and SLy7 forces in Table~\ref{tab:matter}. 
In the table, those for the Skyrme SIII forces are also listed for comparison. 
In the old versions of Skyrme interaction such as the SIII force, the parameters are adjusted to fit the data of some nuclei such as doubly closed-shell nuclei. In the progress of nuclear matter studies, matter properties are used in adjusting parameters of modern interactions as well as the ground state properties. 
For the Skyrme SLy7 force, saturation density $\rho_{\rm 0, eq}$, incompressibility $K_\infty$, equation of state of neutron matter and so on are adopted as the input data for parameter adjusting.\cite{cha98} 
As a result, the SLy7 force gives lower incompressibility $K_\infty$ and effective mass $m^*_0/m$ than the SIII force. 

\section{Deformed structures in $p$- and $sd$-shell region}
\label{sec:deformations}
\subsection{$N=Z=\mbox{even}$ $p$-shell nuclei}
\label{sec:p-shell}
\begin{figure}[btp]
\begin{center}
\begin{tabular}{cc}
\huge (a) $^8$Be & \huge (b) $^{12}$C \\
\includegraphics[width=0.45\textwidth]{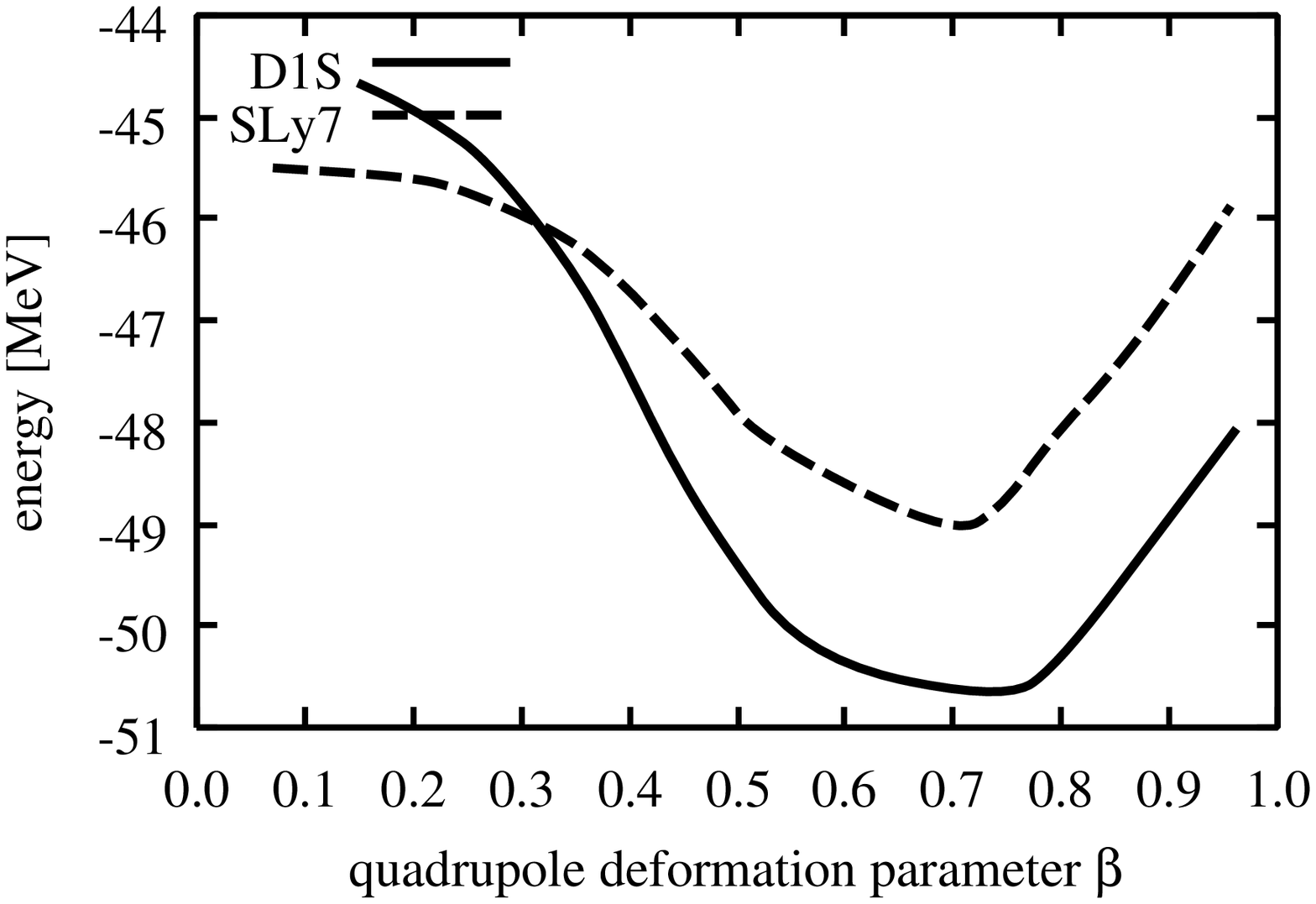} & 
\includegraphics[width=0.45\textwidth]{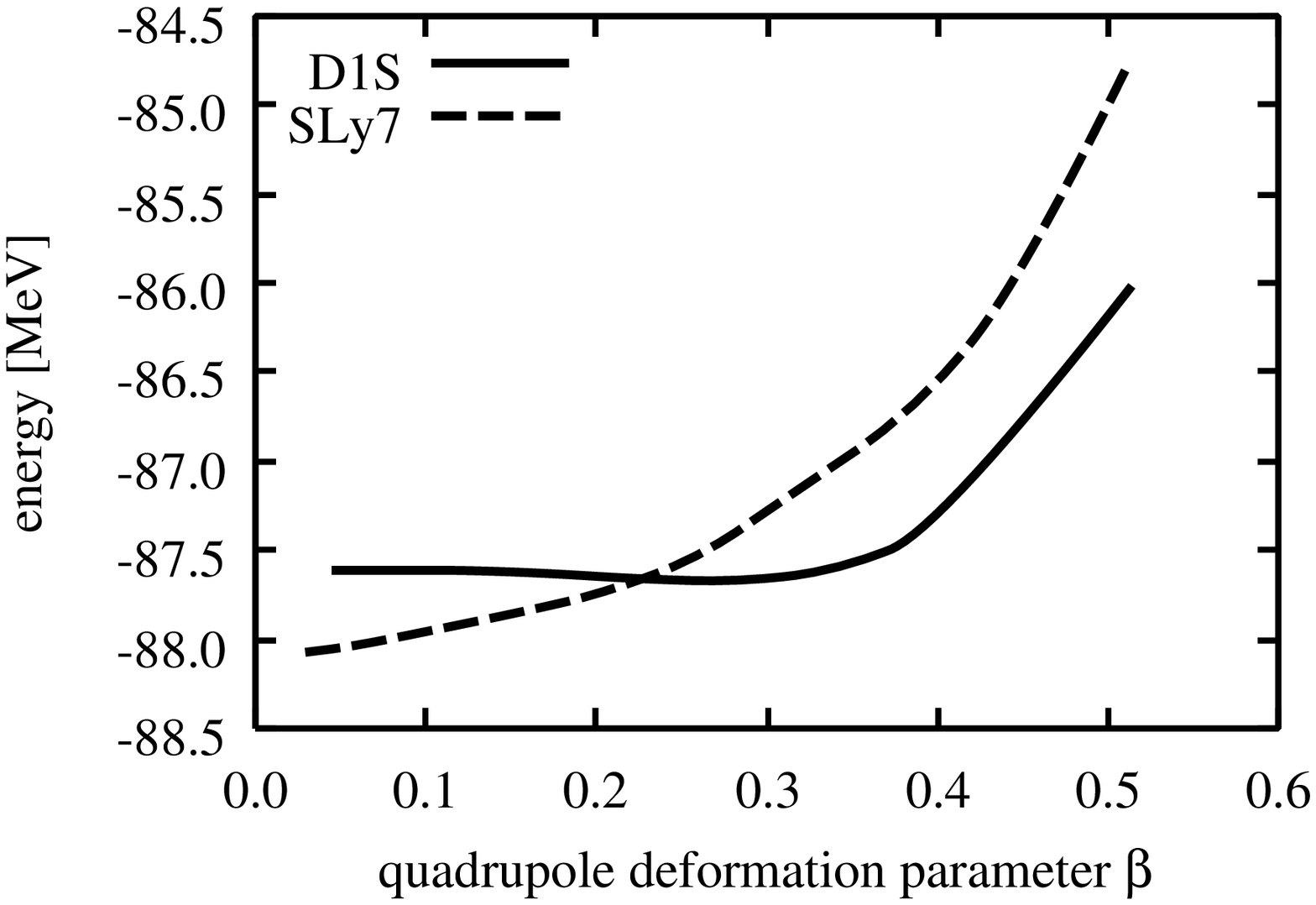} \\
\huge (c) $^{16}$O  \\
\includegraphics[width=0.45\textwidth]{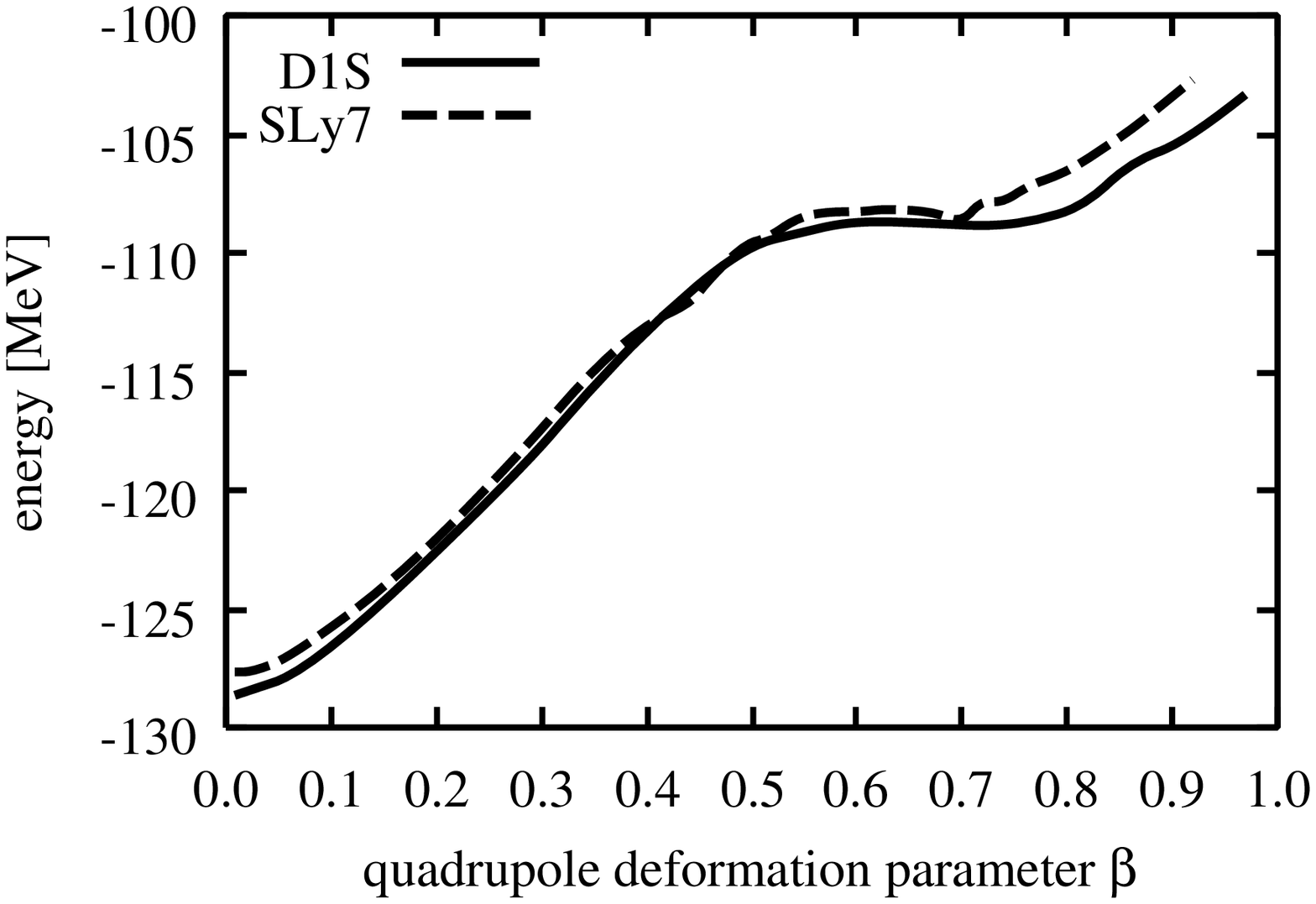}
\end{tabular}
\caption{Energy curves as functions of quadrupole deformation parameter $\beta$ for positive-parity states of (a) $^8$Be, (b) $^{12}$C, and (c) $^{16}$O using the D1S and SLy7 forces. }
\label{fig:p-shell}
\end{center}
\end{figure}
We have performed an energy variation after a projection to a positive-parity state imposing the $\beta$ constraint for $N=Z=\mbox{even}$ $p$-shell nuclei. 
Figure~\ref{fig:p-shell} shows the obtained energy curves as functions of matter quadrupole deformation $\beta$ ($\beta$ energy curves). 

\newpage
\subsubsection{$^8$Be}

\begin{wrapfigure}{l}{0.5\textwidth}
  \begin{center}
    \includegraphics[width=0.4\textwidth]{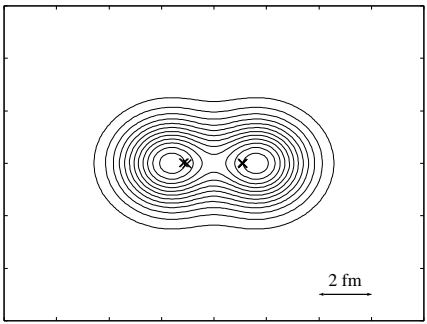}
    \caption{Density distributions at the local minimum on the energy curves ($\beta = 0.75$) for $^8$Be using the D1S force. 
      Symbols ``$\times$'' indicate the centroids of wave packets. }
    \label{fig:8Be_density}
  \end{center}
\end{wrapfigure}

The obtained energy curves for positive-parity states in $^8$Be are shown in Fig.~\ref{fig:p-shell}(a). 
In both cases of the D1S and SLy7 forces, each energy curve has one minimum corresponding to the ground state at $\beta\sim 0.7$. 
In order to analyze the intrinsic structure at the minima, we give the density distribution of the intrinsic wave function in Fig.~\ref{fig:8Be_density}. 
As seen in the figure, $2\alpha$ cluster structure develops in the energy minimum state. 
It is consistent with the well known $2\alpha$ cluster structure in the ground state of $^8$Be, which has been studied by $2\alpha$ cluster models.\cite{hiu72,THSR} 
The distance between two $\alpha$ clusters is approximately 2 fm in the case of the D1S force. 
The binding energy of the ground state in the case of the D1S force is approximately 2 MeV larger than that in the case of the SLy7 force. 

\subsubsection{$^{12}$C}

The energy curves for positive-parity states in $^{12}$C are shown in Fig.~\ref{fig:p-shell}(b). 
In the deformed region, an oblate shape is favored in both cases of the D1S and SLy7 forces, but the shape at the ground states is different between the D1S and SLy7 forces. 
In the case of the D1S force, the oblately deformed state becomes the energy minimum at $\beta = 0.28$, while the SLy7 gives the spherical shape of the energy minimum solution. 
Since $^{12}$C is considered to have the oblate shape in the ground band, the D1S force gives more reasonable result for the deformation of $^{12}$C than the SLy7 within the mean-field approximation. 
However, even in the case of the SLy7 force, the energy curve is quite soft, where the difference of energies at $\beta\sim 0$ and 0.5 is only 3 MeV. 
Therefore, if such treatment as angular momentum projection (AMP) and superposition with respect to $\beta$ are incorporated, the oblate shape wave function may be significantly contained in the ground state. 
In other words, those interactions give different property in the mean-field framework but they may give similar results in a beyond-mean-field framework such as AMP and superposition. 
It is necessary to use beyond-mean-field approaches in order to discuss the detail structures in $^{12}$C where the energy curve is soft along the deformation. 
In large $\beta$ region, $3\alpha$ cluster structures develop. 
We do not present the energy curves in very large $\beta$ region as $\beta > 0.5$, 
because it is difficult to obtain convergent solutions in the region. 

\subsubsection{$^{16}$O}

\begin{figure}[tbp]
  \begin{center}
      \begin{minipage}{0.5\textwidth}
	\includegraphics[width=\textwidth]{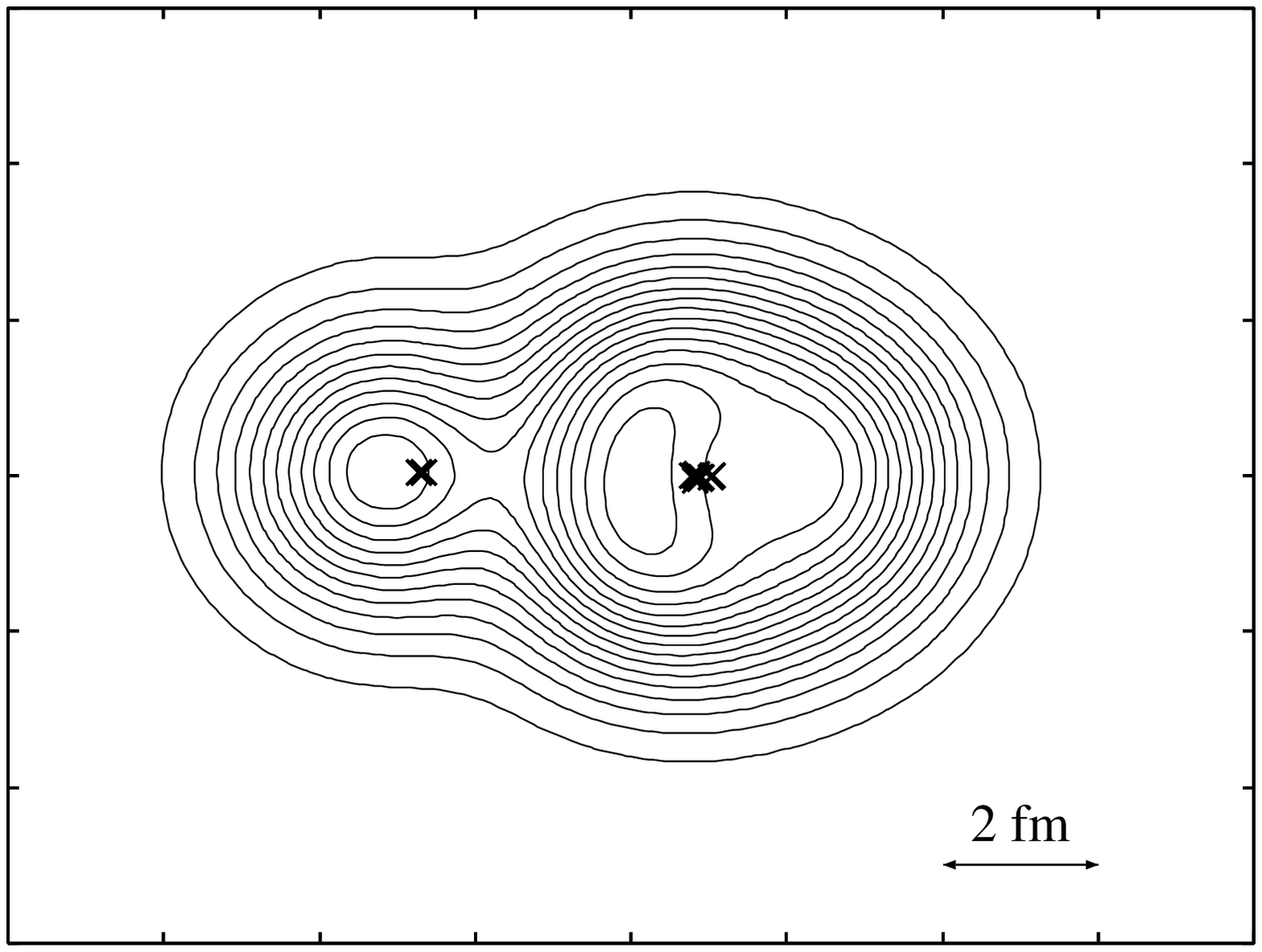} 
	\caption{
	  Density distribution at $\beta=0.60$ for $^{16}$O. 
	  Symbols ``$\times$'' indicate centroids of wave packets. 
	}
	\label{fig:16O_density}
      \end{minipage}
      \begin{minipage}{0.5\textwidth}
	\includegraphics[width=\textwidth]{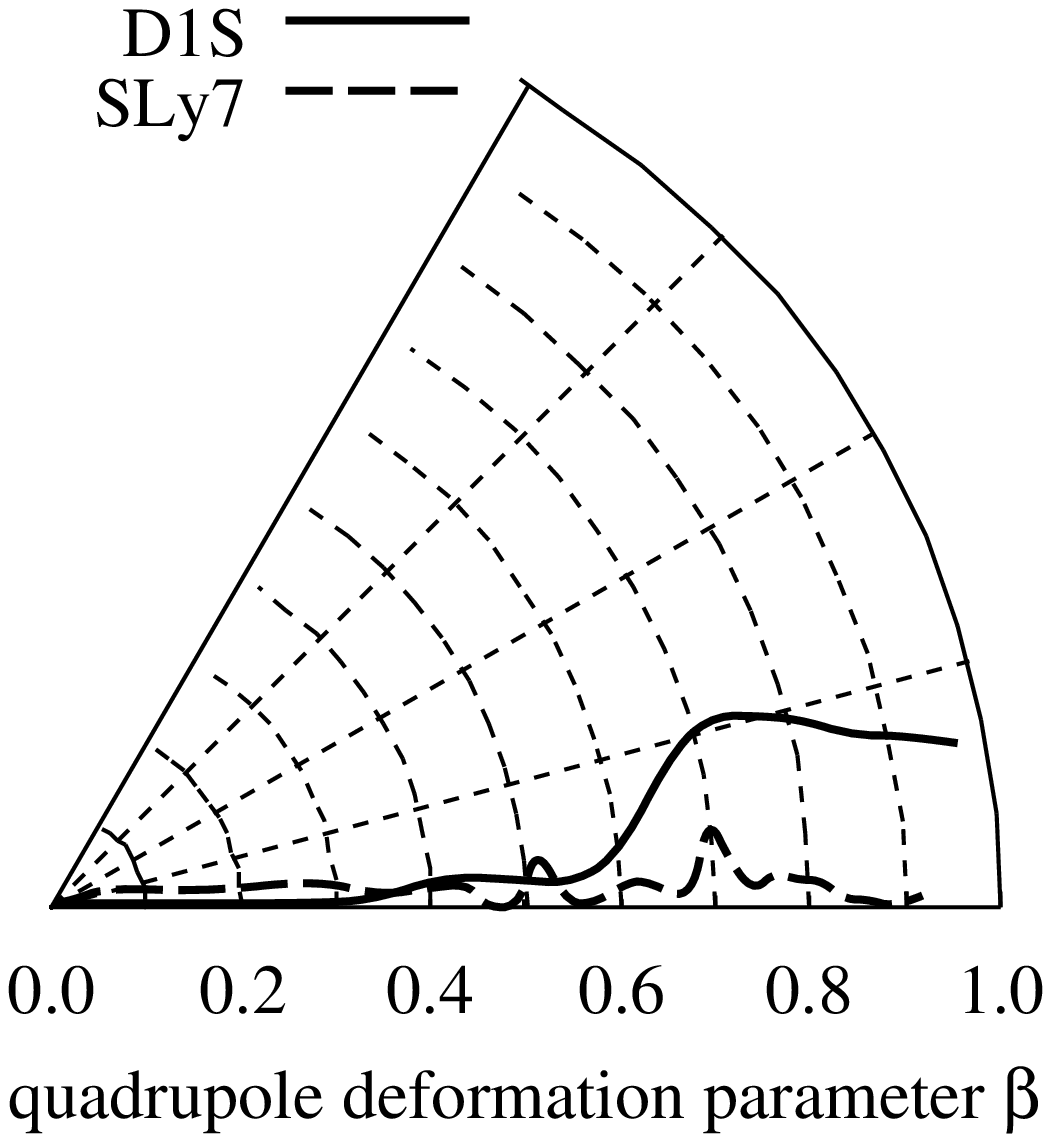}
	\caption{Projection of the energy curves onto the $\beta$-$\gamma$ plane for $^{16}$O. }
	\label{fig:16O_gamma}
      \end{minipage}
  \end{center}
\end{figure}

The energy curves for positive-parity states in $^{16}$O are shown in Fig.~\ref{fig:p-shell}(c). 
The two forces give quantitatively similar $\beta$ energy curves, in each of which the spherical energy minimum for the ground states and a shoulder around $\beta= 0.7$ exist. 
In the intrinsic wave function around the shoulders, we find a developed $\alpha$-$^{12}$C cluster structure as shown in Fig.~\ref{fig:16O_density} for the density distributions. 
It may correspond to the excited band with the developed $\alpha$-$^{12}$C cluster structure which is considered to correspond to the $K^\pi = 0^+$ band built on the $J^\pi = 0_2^+$ state.\cite{hor72,fuj80} 
It is interesting that the $\gamma$ values around the shoulders are different between the results of the D1S and SLy7 forces. 
As shown in Fig.~\ref{fig:16O_gamma}, triaxially deformed states are obtained in $\beta \gtrsim 0.6$ region in the case of the D1S force, while axial symmetric shapes are obtained in all region in the case of the SLy7 force. 
It comes from the shapes of $^{12}$C clusters composing the $\alpha$-$^{12}$C cluster structure. 
As mentioned before, a $^{12}$C system has oblate and spherical shapes at the energy minimal states in the cases of the D1S and SLy7 forces, respectively. 
This trend is found also in the excited states of $^{16}$O with the $\alpha$-$^{12}$C cluster. 
In fact, the deformations of $^{12}$C clusters are oblate and spherical in the cases of the D1S and SLy7 forces, respectively. 
As a result, an axial symmetric $\alpha$-$^{12}$C cluster structure is obtained because of spherical $^{12}$C cluster in the case of the SLy7 force. 
In the case of the D1S force, $\alpha$ cluster locate on the edge of the oblate $^{12}$C cluster, and total system forms the triaxial shape. 

Compared to SHF + BCS assuming reflection symmetry using the Skyrme SLy4 (SLy4) force\cite{ben03b}, the excitation energy around the shoulder at $\beta\sim 0.5$--0.7 is approximately 13 MeV higher, though energy variation after parity projection are performed in present study. 
It is unnatural because the SHF + BCS assuming reflection symmetry can not describe an $\alpha$-$^{12}$C cluster state, which is well known structure in excited states and expected to be appear in deformed region. 
The low excitation energy in the SHF + BCS may be owing to very strong pairing potential. 

\subsection{$N=Z=\mbox{even}$ $sd$-shell nuclei}
\label{sec:sd-shell}

\begin{figure}[tbp]
\begin{center}
\begin{tabular}{ll}
\huge (a) $^{20}$Ne (positive) & \huge (b) $^{20}$Ne (negative) \\
\includegraphics[width=0.45\textwidth]{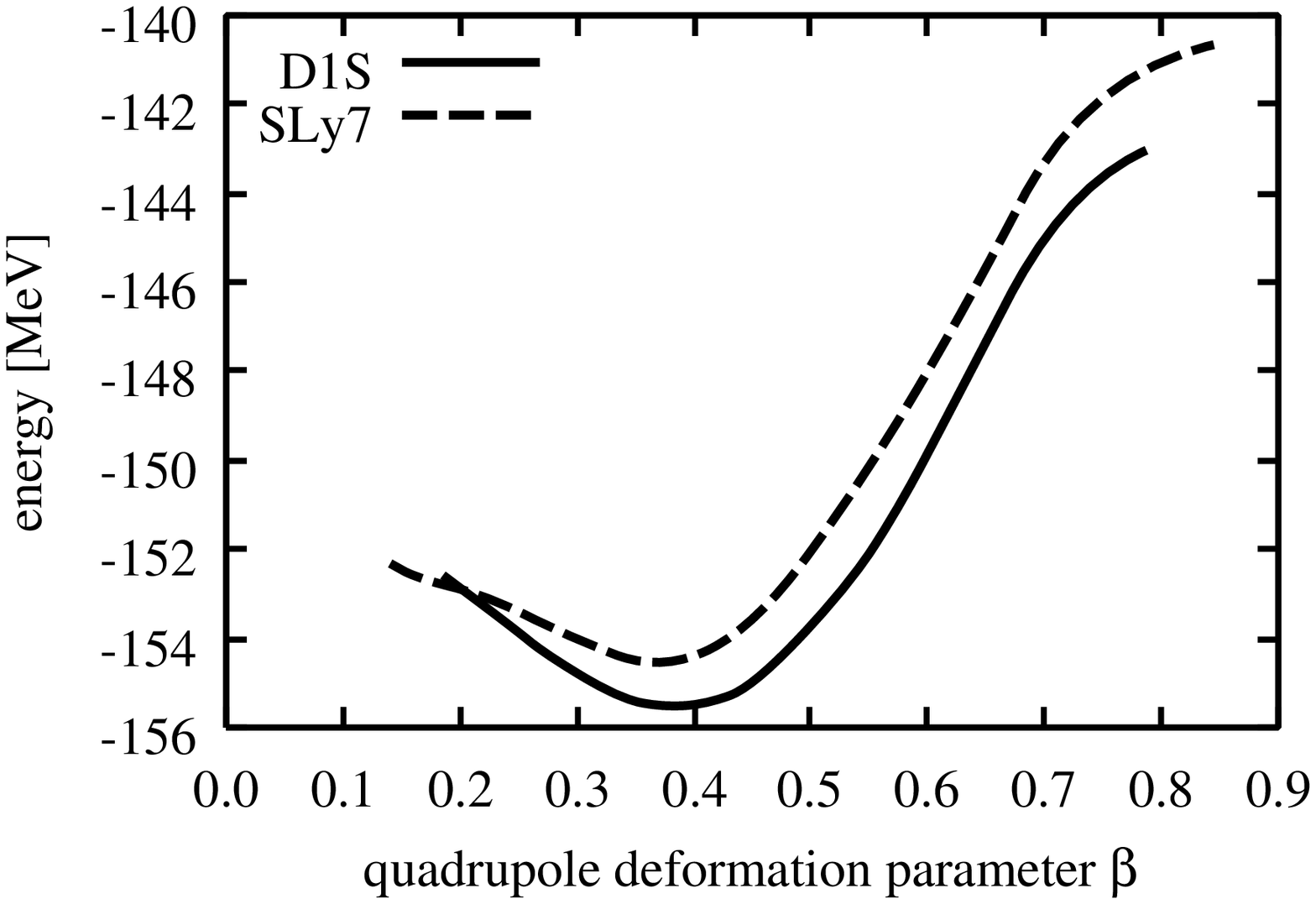} &
\includegraphics[width=0.45\textwidth]{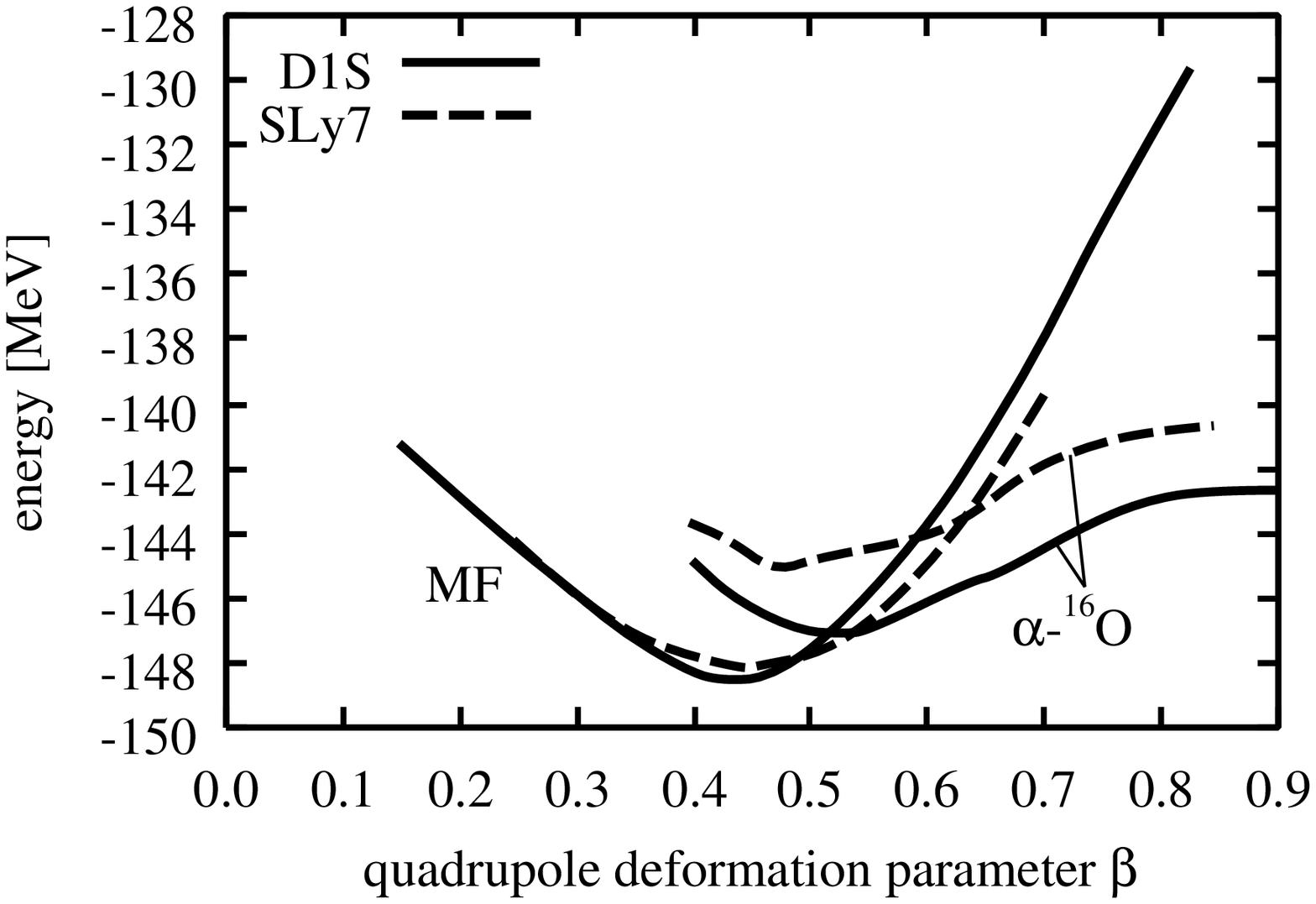} \\\\
\huge (c) $^{24}$Mg (positive) & \huge (d) $^{28}$Si (positive) \\
\includegraphics[width=0.45\textwidth]{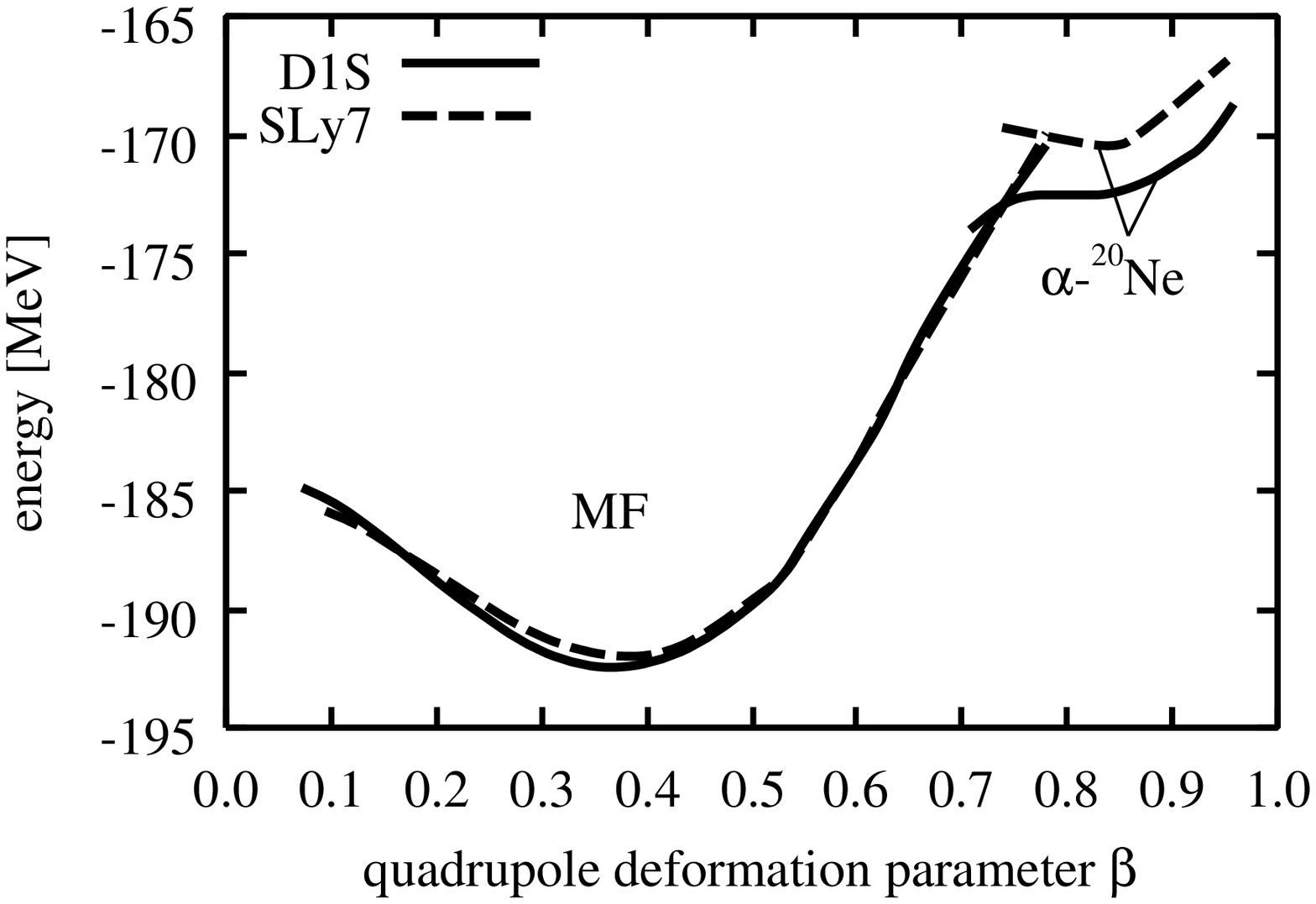} &
\includegraphics[width=0.45\textwidth]{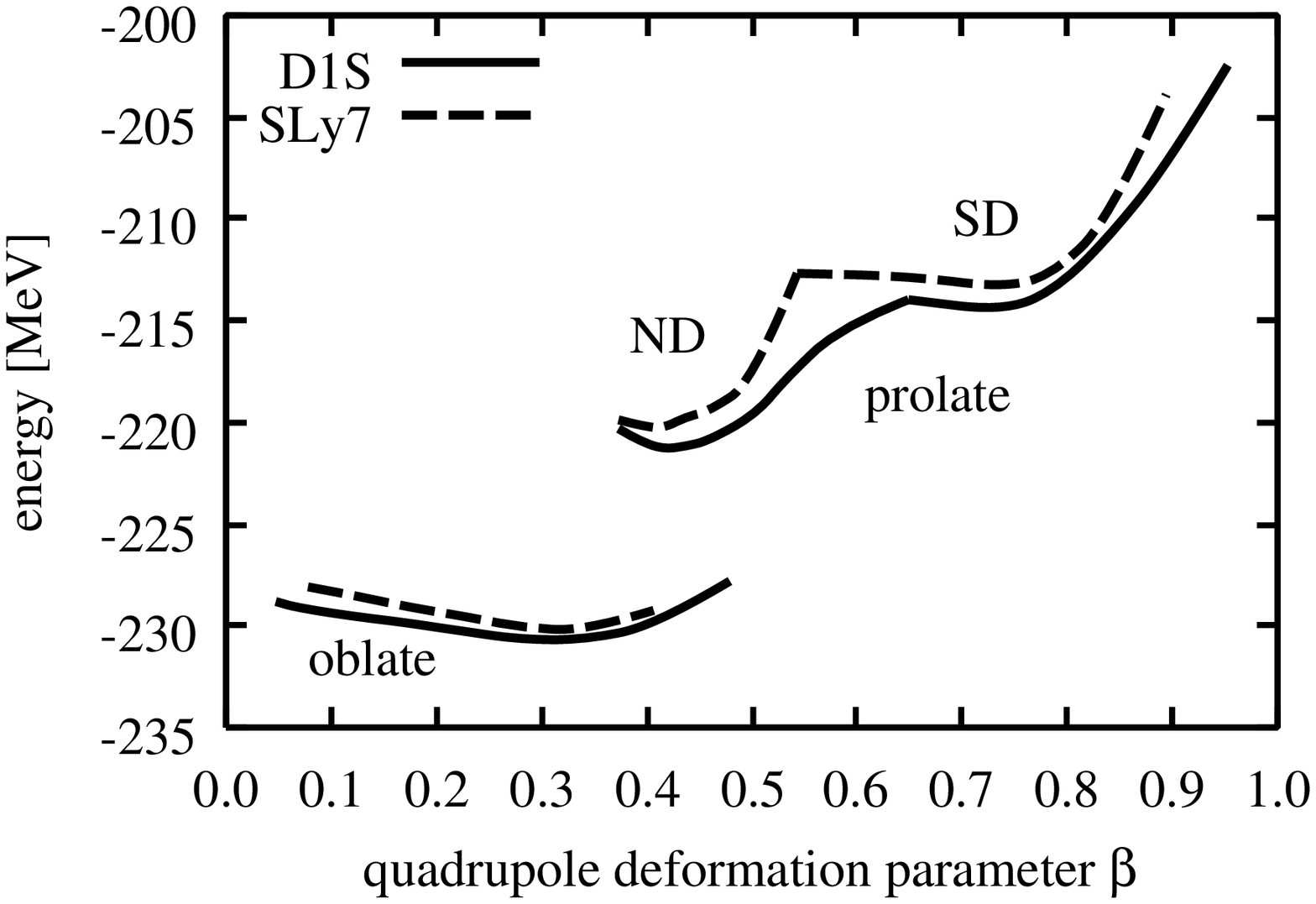} \\\\
\huge (e) $^{32}$S (positive) & \huge (f) $^{36}$Ar (positive) \\
\includegraphics[width=0.45\textwidth]{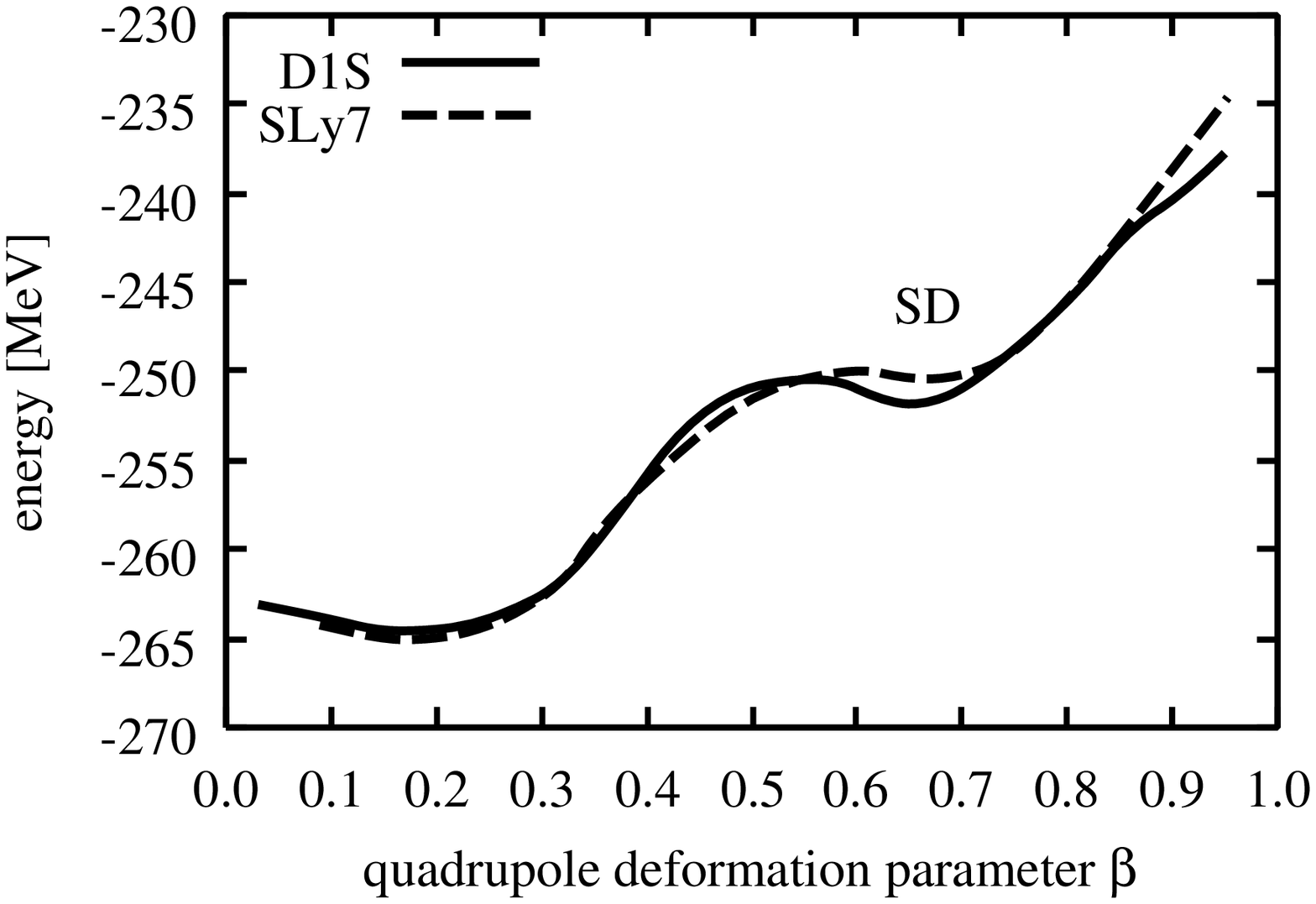} &
\includegraphics[width=0.45\textwidth]{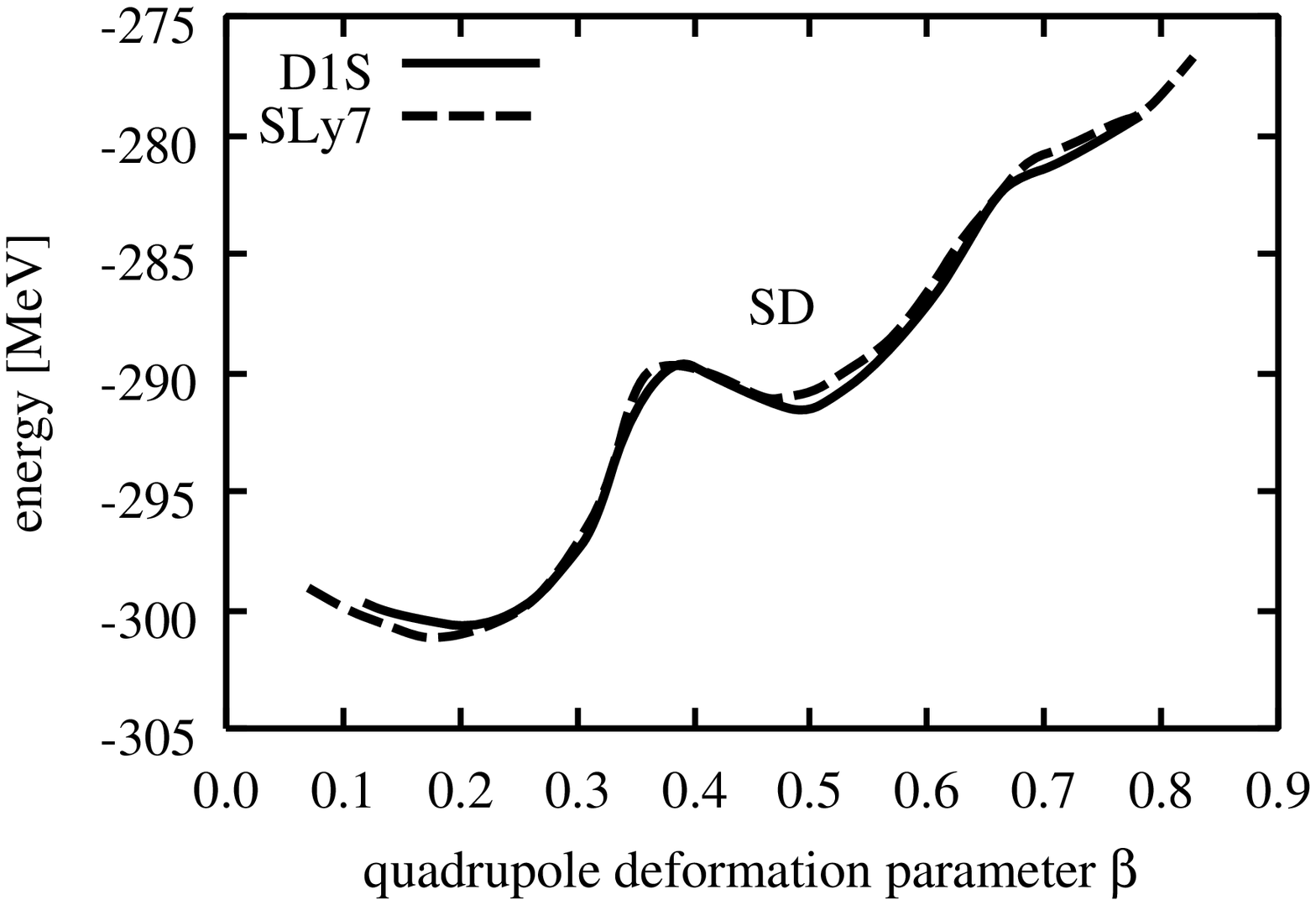} 
\end{tabular}
\caption{Energy curves as functions of quadrupole deformation parameter $\beta$ for (a) and (b) $^{20}$Ne, (c) $^{24}$Mg, (d) $^{28}$Si, (e) $^{32}$S, and (f) $^{36}$Ar using the D1S and SLy7 force. 
Figure (b) are the energy curves for negative-parity states, and others are for positive-parity states. 
``MF'' indicates ``mean-field-type structure'', and ``ND'' and ``SD'' indicate ``normal-deformed'' and ``superdeformed'', respectively (see text). 
}
\label{fig:sd-shell}
\end{center}
\end{figure}

We have obtained $\beta$ energy curves by an energy variation after a projection to positive-parity states imposing $\beta$ constraint in $sd$-shell $N=Z={\rm even}$ nuclei. 
We have also performed an energy variation for negative-parity states of $^{20}$Ne which has the well known $\alpha$-$^{16}$O cluster structure. 
Figures~\ref{fig:sd-shell} show the energy curves for the $N=Z=\mbox{even}$ $sd$-shell nuclei except for $^{40}$Ca. 
The detailed study of structures in $^{40}$Ca using the D1S and SLy7 forces was done in Ref.~\citen{tan07}. 

\newpage

\subsubsection{$^{20}$Ne}
\begin{figure}[tbp]
\begin{center}
\begin{tabular}{ll}
\huge (a) & \huge (b) \\
\includegraphics[width=0.4\textwidth]{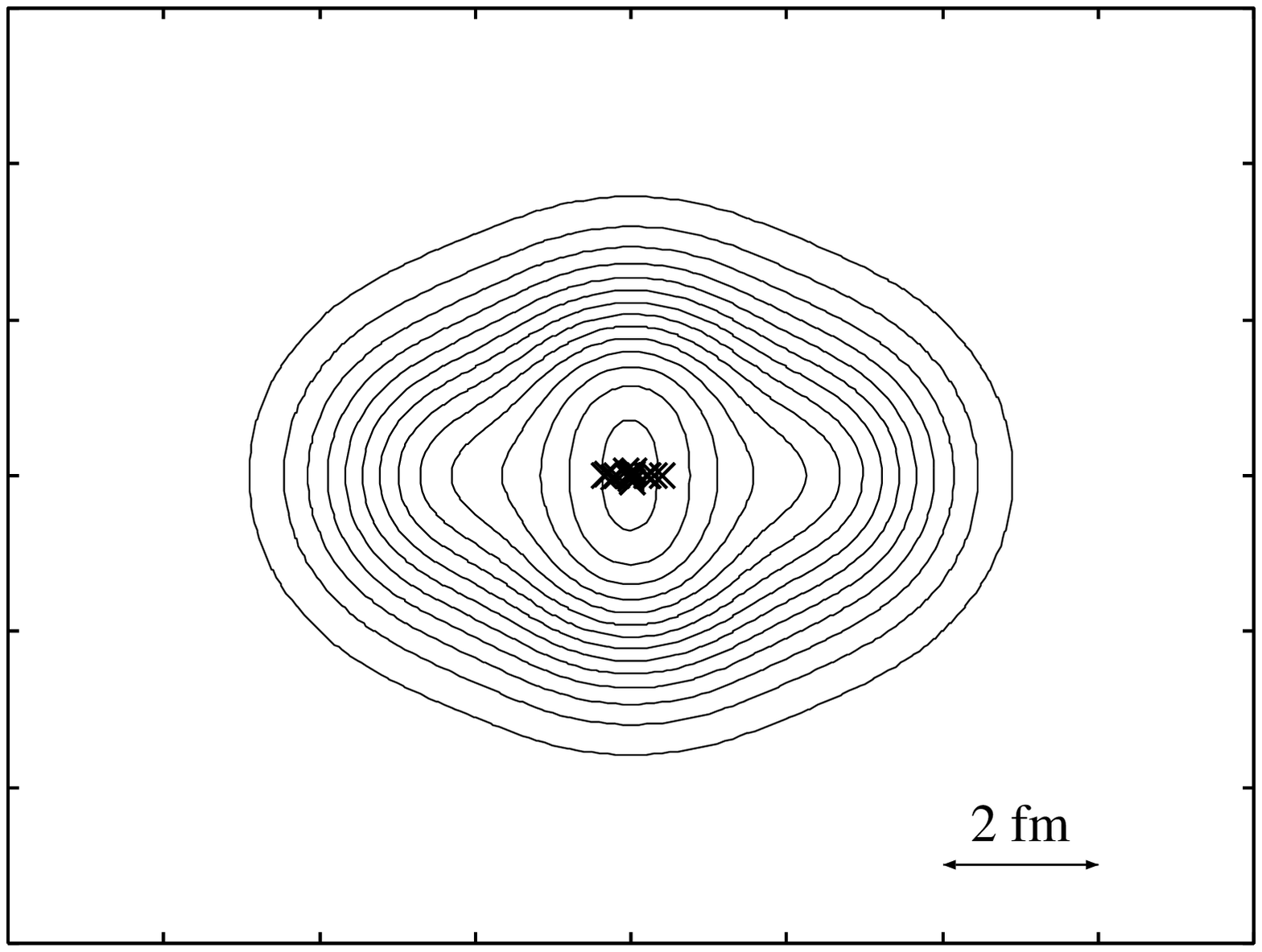} &
\includegraphics[width=0.4\textwidth]{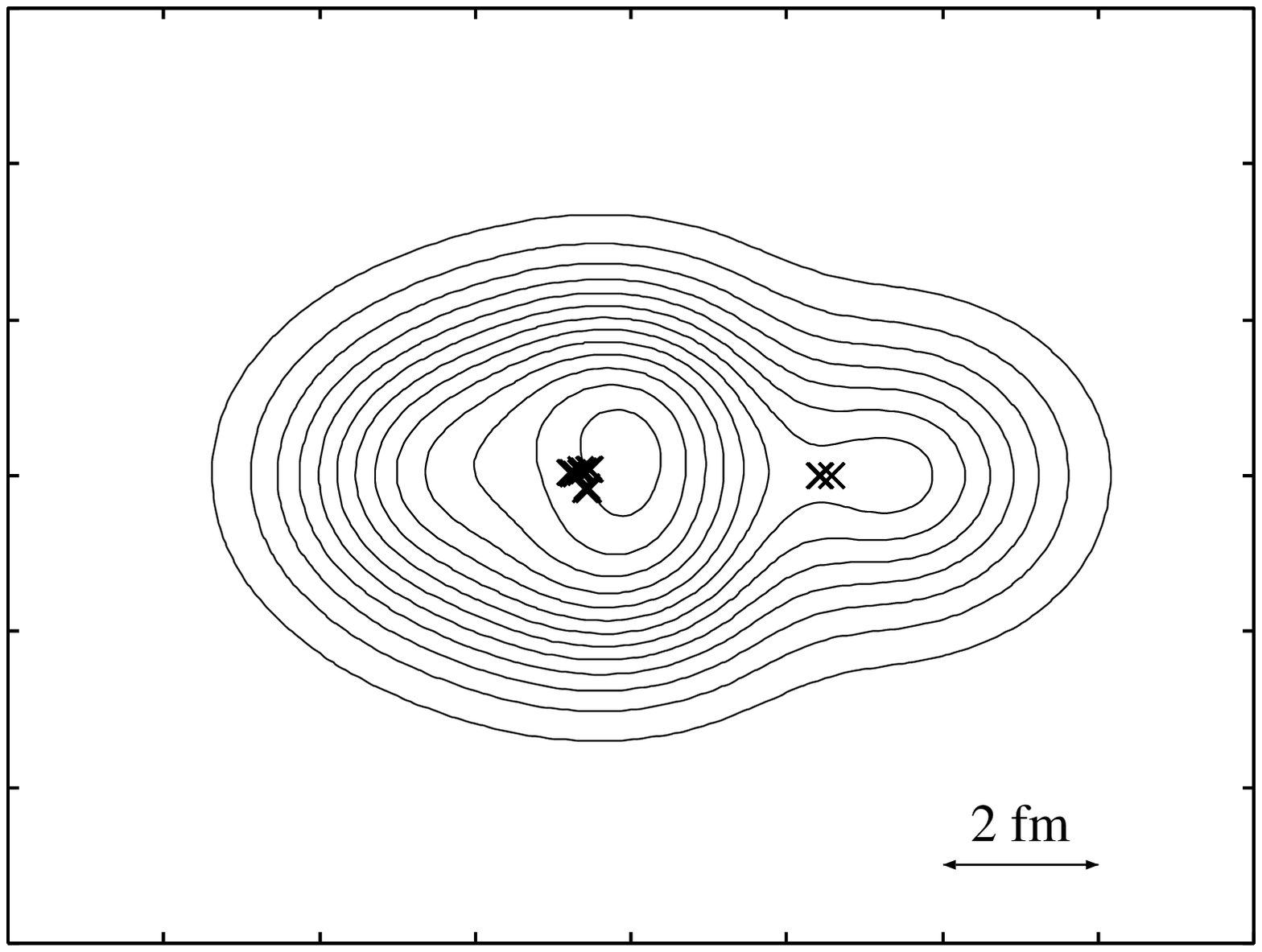}\\
\\
\huge (c) & \huge (d) \\
\includegraphics[width=0.4\textwidth]{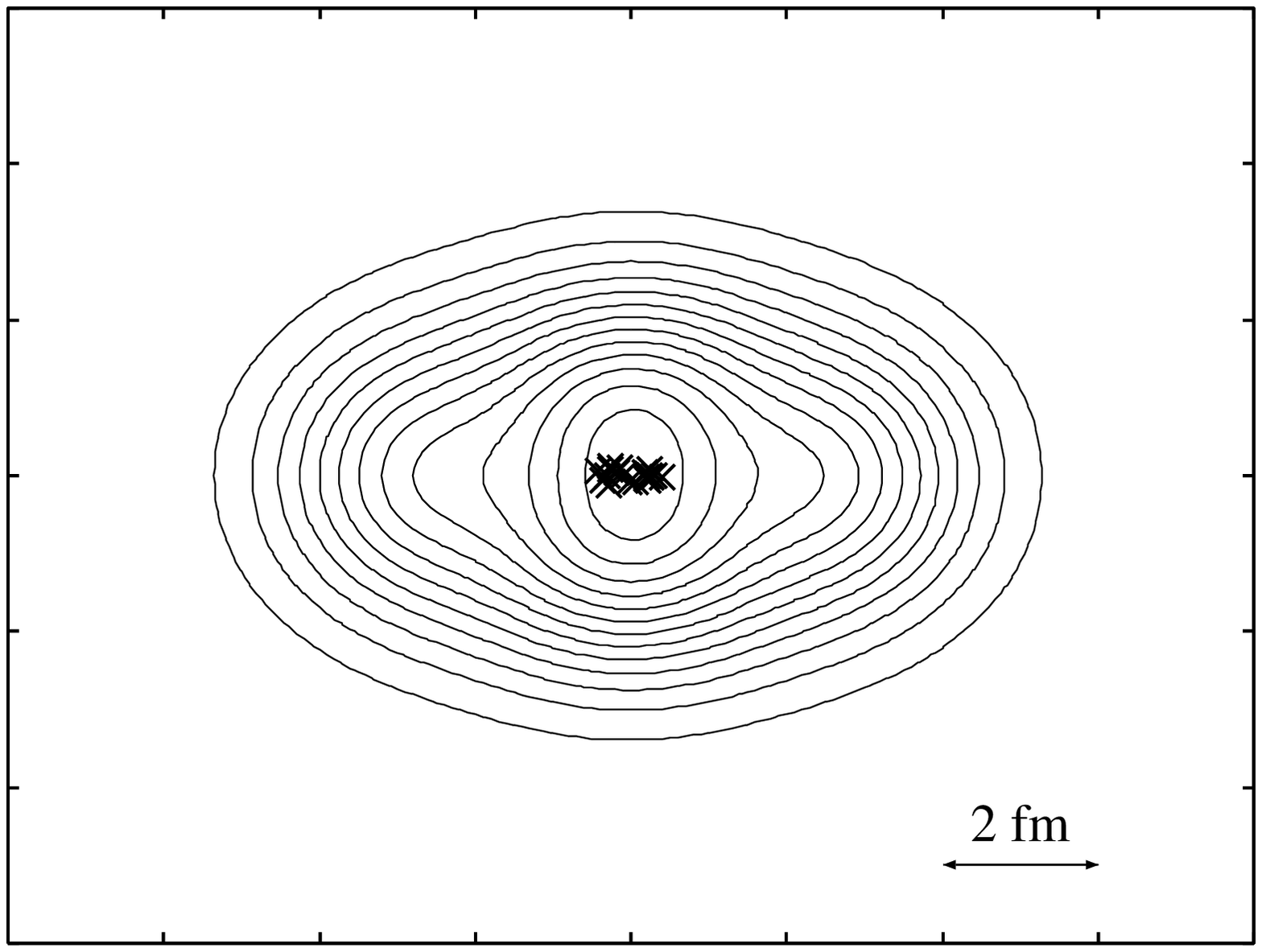} &
\includegraphics[width=0.4\textwidth]{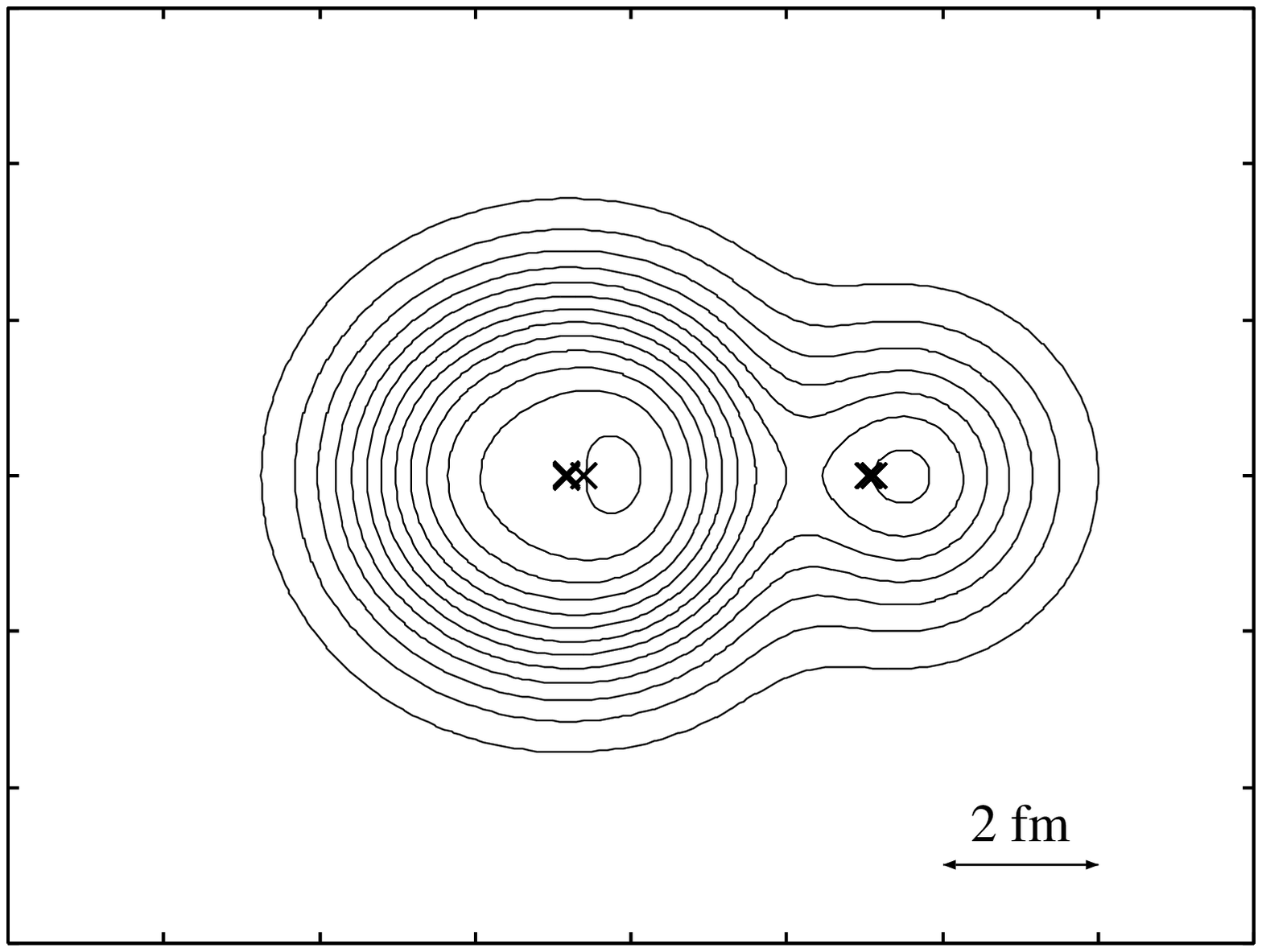}
\end{tabular}
\caption{
Density distributions of (a) the ground state ($\beta=0.37$), (b) a deformed state in positive-parity state ($\beta=0.62$), (c) a mean-field-type structure in negative-parity states at $\beta=0.52$, and (d) an $\alpha$-$^{16}$O cluster structure in negative-parity states at $\beta=0.53$ for $^{20}$Ne using the D1S force. 
Symbols ``$\times$'' indicate centroids of wave packets. 
}
\label{fig:20Ne_density}
\end{center}
\end{figure}

\begin{wrapfigure}{l}{0.5\textwidth}
  \begin{center}
    \includegraphics[width=0.45\textwidth]{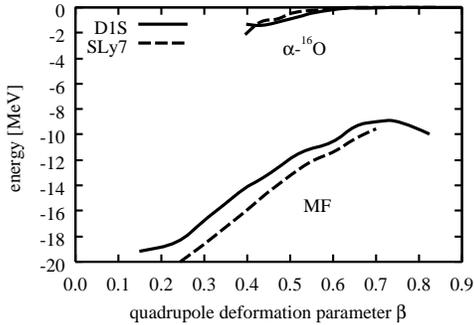}
    \caption{
      Spin-orbit energy curves for negative-parity states of $^{20}$Ne. 
      The curves labeled ``MF'' and ``$\alpha$-$^{16}$O'' are for mean-field-type and an $\alpha$-$^{16}$O cluster structures, respectively. 
    }
    \label{fig:20Ne_LS-}
  \end{center}
\end{wrapfigure}

The energy curves for positive-parity states of $^{20}$Ne are shown in Fig.~\ref{fig:sd-shell}(a). 
The two forces give qualitatively similar curves that have prolate minima corresponding to the ground states at $\beta\sim 0.4$. 
Around the minima, developed cluster structure is not seen, 
but, with the increase of the quadrupole deformation parameter $\beta$, $\alpha$-$^{16}$O clustering is developing. 
Density distributions of the minimum ($\beta = 0.37$) and largely deformed state ($\beta=0.62$) in the case of the D1S force are shown in Fig.~\ref{fig:20Ne_density}(a) and (b), respectively. 
The binding energy in the case of the D1S force is larger than those in the case of the SLy7 force. 
Compared to HFB using the D1S force,\cite{rod03} the binding energy of the ground state is approximately 3 MeV larger. 
More detailed discussions of positive-parity states in $^{20}$Ne are presented in Ref.~\citen{kim04a} and \citen{tan04}. 

The energy curves for negative-parity states in $^{20}$Ne are shown in Fig.~\ref{fig:sd-shell}(b). 
In the energy variation for a given constraint value of $\beta$, we have obtained two different types of structure. 
One is the deformed mean-field-type structure with no developed cluster (``MF'') and the other is $\alpha$-$^{16}$O cluster structure (``$\alpha$-$^{16}$O''). 
In the small $\beta$ region, the energy of MF state is lower than the $\alpha$-$^{16}$O state, while, in the largely deformed region, that of the $\alpha$-$^{16}$O state becomes lower. 
The density distributions at the crossing point of the energy curves labeled ``MF'' and ``$\alpha$-$^{16}$O'' in the case of the D1S force are shown in Figs.~\ref{fig:20Ne_density}(c) and (d), respectively. 
Expectation values of the spin-orbit term for $\alpha$-$^{16}$O cluster structure are much smaller in magnitude than those for mean-field-type structures as shown in Fig.~\ref{fig:20Ne_LS-}, because the expectation values of the spin-orbit term of the ground states in $\alpha$ and $^{16}$O are almost zero. 
The mean-field-type and $\alpha$-$^{16}$O cluster structures that appear on the curves should correspond to the $K^\pi = 2^-$ band built on the $J^\pi = 2_1^-$ state and the  $K^\pi = 0^-$ band built on the $J^\pi = 1_2^-$ state, respectively.\cite{kim04a} 
Comparing the energies of these states between the D1S and SLy7 forces, we have found that the energy of the $\alpha$-$^{16}$O cluster structure in the case of the D1S force is approximately 2 MeV lower than that in the case of the SLy7 force, 
whereas the energies of mean-field-type structures are quite similar to each other. 
It suggests that the SLy7 force gives higher excitation energy of the $\alpha$-$^{16}$O cluster state than that of the D1S force. 

\subsubsection{$^{24}$Mg}

\begin{wrapfigure}{l}{0.5\textwidth}
  \begin{center}
    \includegraphics[width=0.4\textwidth]{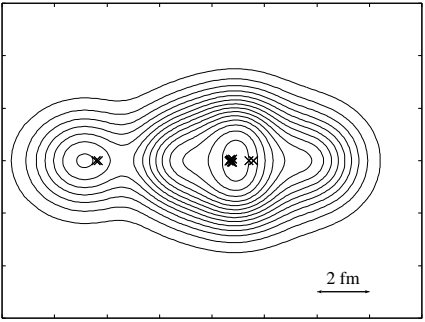}
    \caption{Density distribution of $\alpha$-$^{20}$Ne cluster structure in $^{24}$Mg ($\beta=0.84$) using the D1S force. }
    \label{fig:24Mg_density}
  \end{center}
\end{wrapfigure}

The energy curves for positive-parity states in $^{24}$Mg are shown in Fig.~\ref{fig:sd-shell}(c). 
The two forces give qualitatively similar energy curves. 
The mean-field-type structure (``MF'') and the developed $\alpha$-$^{20}$Ne cluster structure (``$\alpha$-$^{20}$Ne'') are obtained in small and large $\beta$ region, respectively. 
The curves for the mean-field-type structure is quite similar in both cases of the D1S and SLy7 forces, while the energies of the $\alpha$-$^{20}$Ne cluster structures for the D1S force are approximately 2 MeV lower than those for the SLy7 force. 
The mean-field-type and $\alpha$-$^{20}$Ne cluster structures have minima are at $\beta\sim 0.4$ and 0.8, respectively. 
Both are prolate deformation. 
Excitation energies of the local minima in the curves of the $\alpha$-$^{20}$Ne cluster structures for the D1S and SLy7 forces are 19.9 and 21.5 MeV, respectively. 
The density distribution of $\alpha$-$^{20}$Ne cluster structure at the local minimum for the D1S force ($\beta=0.84$) is shown in Fig.~\ref{fig:24Mg_density}. 
The $\alpha$ cluster locates on the axial symmetric axis of a $^{20}$Ne cluster. 
This $\alpha$-$^{20}$Ne cluster structure with the large prolate deformation is described by four particle excitation from $sd$-shell to $pf$-shell in the shell-model representation. 
This is consistent with HFB calculation using the D1S force.\cite{egi04} 
Compared to HFB using the D1S force\cite{rod02}, the binding energy of the ground state is approximately 3 MeV larger. 

\subsubsection{$^{28}$Si}

\begin{figure}[tbp]
  \begin{minipage}{0.5\textwidth}
    \begin{center}
      \includegraphics[width=0.9\textwidth]{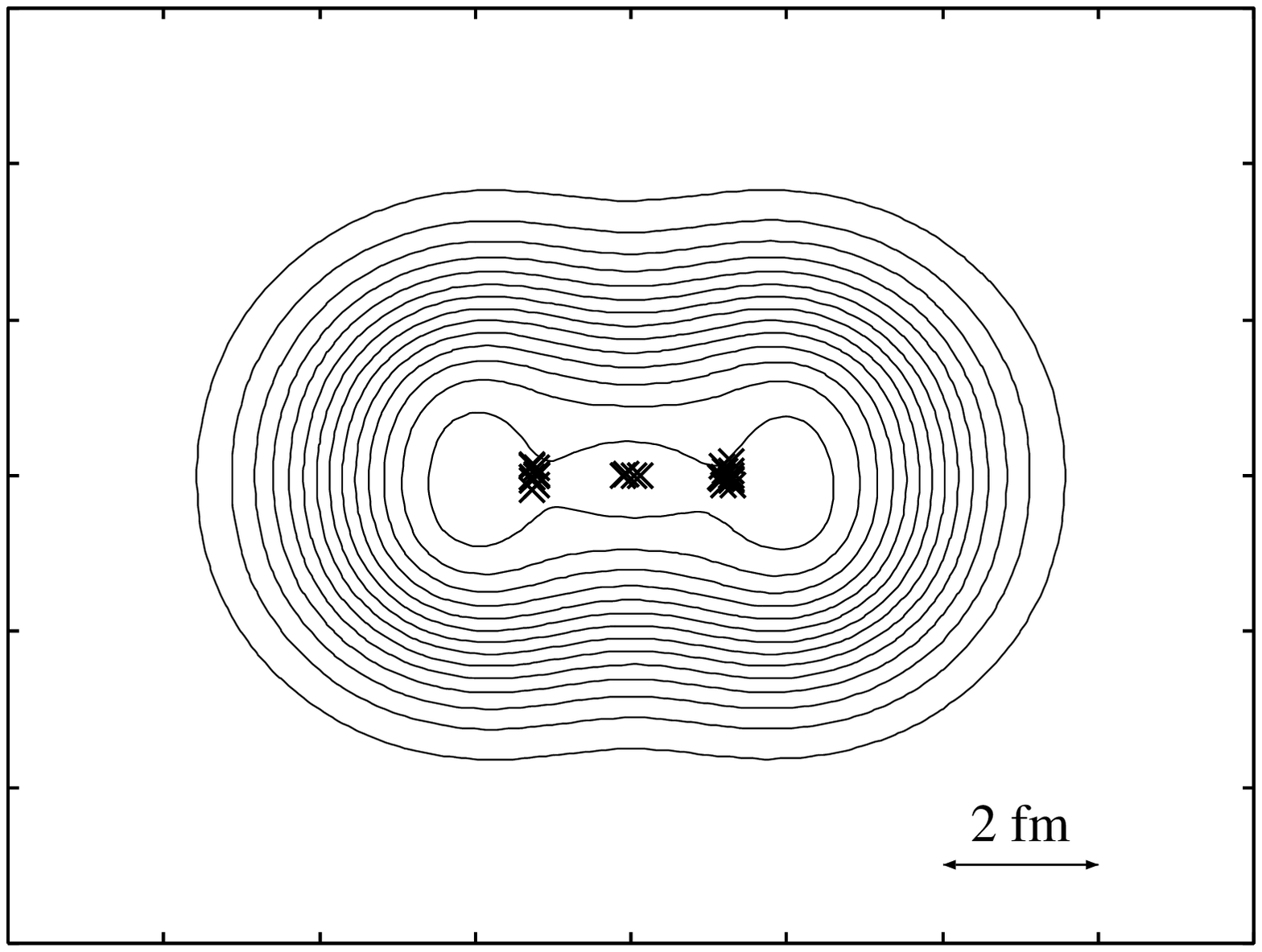}
      \caption{Density distribution of the prolate ND state in $^{28}$Si for the D1S force. 
	Symbols ``$\times$'' indicate centroids of wave packets. }
      \label{fig:28Si_SD}
    \end{center}
  \end{minipage}
  \begin{minipage}{0.5\textwidth}
    \begin{center}
      \includegraphics[width=0.9\textwidth]{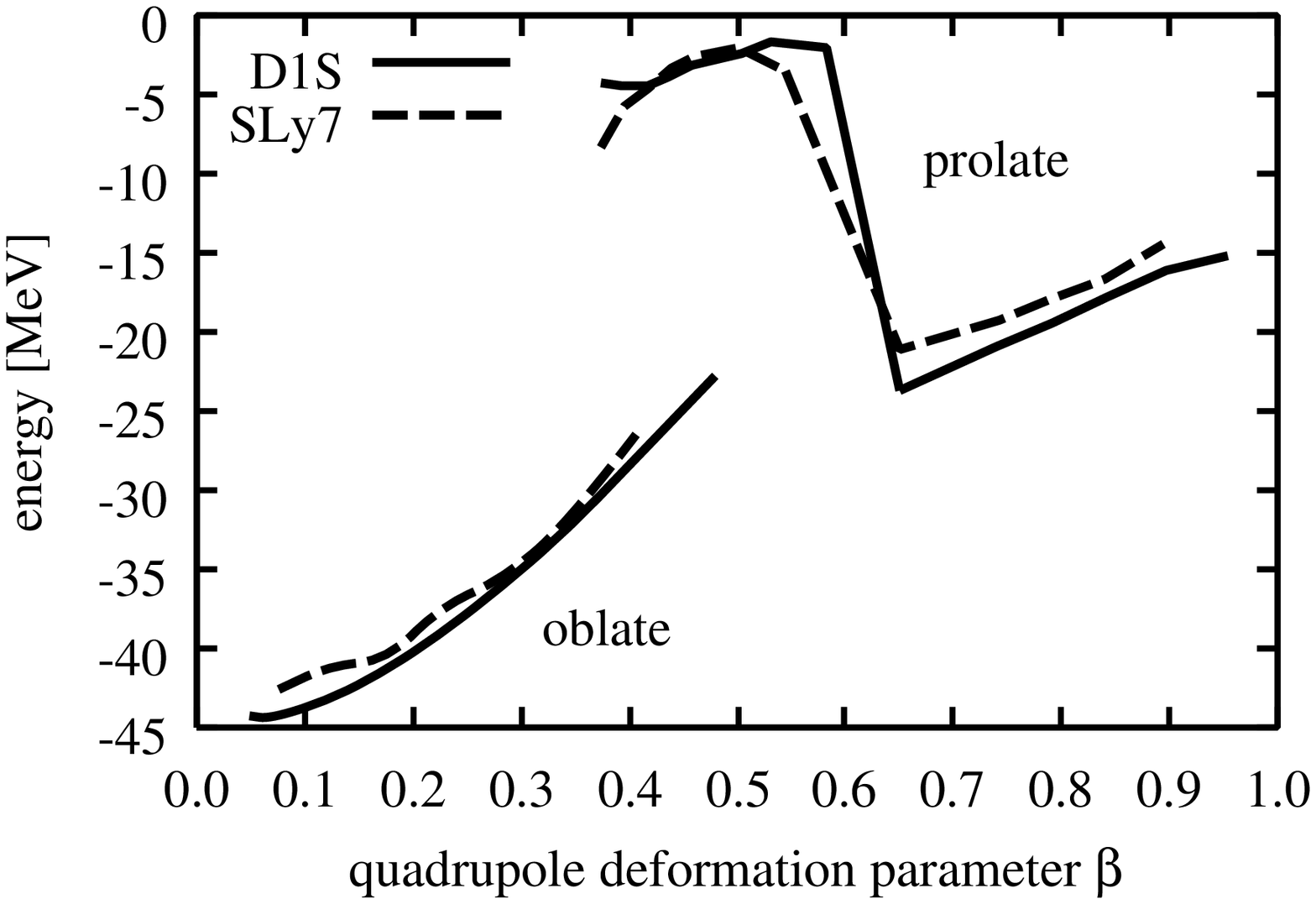}
      \caption{The spin-orbit energy curves for positive-parity states in $^{28}$Si. }
      \label{fig:28Si_LS+}
    \end{center}
  \end{minipage}
\end{figure}

The energy curves for positive-parity states in $^{28}$Si are shown in Fig.~\ref{fig:sd-shell}(d). 
The both forces give quantitatively similar energy curves. 
The curves for prolate and oblate shapes as a function of $\beta$ are obtained. 
The minimum obtained at $\beta\sim0.3$ on the curves for the oblate shape corresponds to the ground state. 
Since the curve for oblate shape  is quite soft against quadrupole deformation, the wave function in the wide $\beta$ range from spherical to oblate are expected to mix into the ground state wave function of $^{28}$Si. 
The curve for prolate shape has two local minima at $\beta\sim 0.4$ and 0.8. 
We call those prolate ND and SD local minima, respectively. 
It is considered that the oblate (ground) and prolate band coexist in $^{28}$Si\cite{28Si}, which is consistent with the present result. 
The excitation energies of the ND and SD minima are estimated to be 9.5 and 16.1 MeV, respectively, of the relative energy to the ground state in the case of the D1S force. 
The prolate ND state in the case of the D1S force is slightly softer against quadrupole deformation than that in the case of the SLy7 force, 
whereas the curves are quantitatively similar around the SD minima in both cases of the D1S and SLy7 forces. 
Around the prolate ND minima, $^{12}$C-$\alpha$-$^{12}$C-like core structures are seen (Fig.~\ref{fig:28Si_SD}), while no cluster structure develops around the minima corresponding to the ground state and the SD minima. 
The expectation values of spin-orbit term are shown in Fig.~\ref{fig:28Si_LS+}. 
It is found that the energy contribution of the spin-orbit term is remarkably small in the prolate ND state. 
It indicates that $^{12}$C-like core in this state are not the $0p_{3/2}$ sub-shell closure but they have the spin saturated structure like a $3\alpha$ structure. 
In order to discuss the details of the coexisting various deformations in $^{28}$Si, it is important to incorporate the shape mixing in beyond-mean-field approaches. 

\subsubsection{$^{32}$S}

\begin{wrapfigure}{l}{0.5\textwidth}
  \begin{center}
    \includegraphics[width=0.45\textwidth]{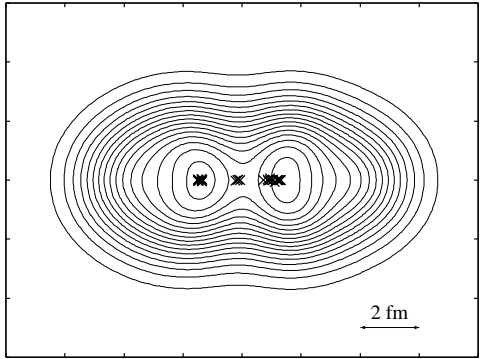}
    \caption{
      Density distribution of the SD state ($\beta=0.67$) in $^{32}$S using the D1S force. 
    }
    \label{fig:32S_SD}
  \end{center}
\end{wrapfigure}

The energy curves for positive-parity states in $^{32}$S are shown in Fig.~\ref{fig:sd-shell}(e). 
The both forces give quantitatively similar energy curves that have the prolate ground states at $\beta\sim 0.2$ and prolate local minima at $\beta\sim 0.7$ (the SD local minima), respectively. 
Excitation energies of the SD local minima are estimated to be 12.7 MeV and 14.6 MeV for the D1S and SLy7 forces, respectively. 
The density distribution around the SD local minimum has a neck structure as shown in Fig.~\ref{fig:32S_SD}. 
It corresponds to $^{16}$O-$^{16}$O cluster structure of the SD state in $^{32}$S.\cite{kim04b} 
Compared to HFB calculation using the D1S,\cite{rod00} the binding energies at the ground state and the SD minimum for the D1S is 4 MeV and 1 MeV larger, respectively. 
The excitation energies at the SD minimum for the SLy7 are a few MeV higher than those with SHF for the SLy4.\cite{ina02} 
The HF + BCS using the Skyrme SLy6 (SLy6) force gives no SD minimum.\cite{ben03a} 

\subsubsection{$^{36}$Ar}

\begin{wrapfigure}{l}{0.5\textwidth}
  \begin{center}
    \includegraphics[width=0.4\textwidth]{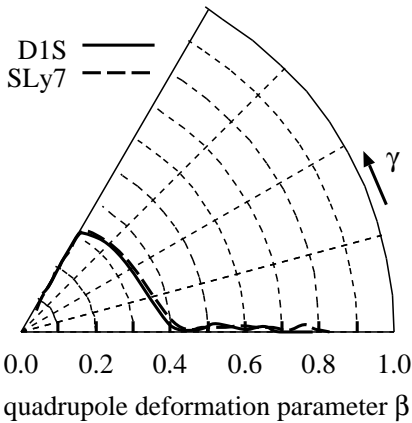}
    \caption{Energy curves projected onto $\beta$-$\gamma$ plane of $^{36}$Ar. }
    \label{fig:36Ar_gamma}
  \end{center}
\end{wrapfigure}

The energy curves for positive-parity states in $^{36}$Ar are shown in Fig.~\ref{fig:sd-shell}(f). 
Both the forces give quantitatively similar energy curves that have minima corresponding to the oblate ground state at $\beta\sim 0.2$ and prolate local minima at $\beta\sim 0.5$ (the SD local minima). 
The $\gamma$ values are changing as increasing $\beta$ values as shown in Fig.~\ref{fig:36Ar_gamma}. 
In the small $\beta$ region, the system has an oblate shape. 
Then, across the local maxima at $\beta\sim 0.35$, the system rapidly changes to prolate shape. 
The excitation energies of the SD local minima in the cases of the D1S and SLy7 forces are 9.2 and 10.0 MeV, respectively. 
The $\beta$ values of the SD minima are consistent with the calculations SHF\cite{ina02} and HFB\cite{rod04}. 
The energies at the SD local minimum in the present calculations for the D1S and SLy7 forces are almost consistent with that of the the HFB using the D1S and that of the SHF using the SLy4 force. 
The HF + BCS using the SLy6 force gives a spherical ground state and no SD minimum.\cite{ben03a} 
Developed cluster structures do not appear in all $\beta$ region. 

\subsection{Neutron-rich nuclei $^{24}$O}
\label{sec:24O}

\begin{figure}[tbp]
  \begin{center}
    \begin{tabular}{cc}
      \huge (a) & \huge (b) \\
      \includegraphics[width=0.5\textwidth]{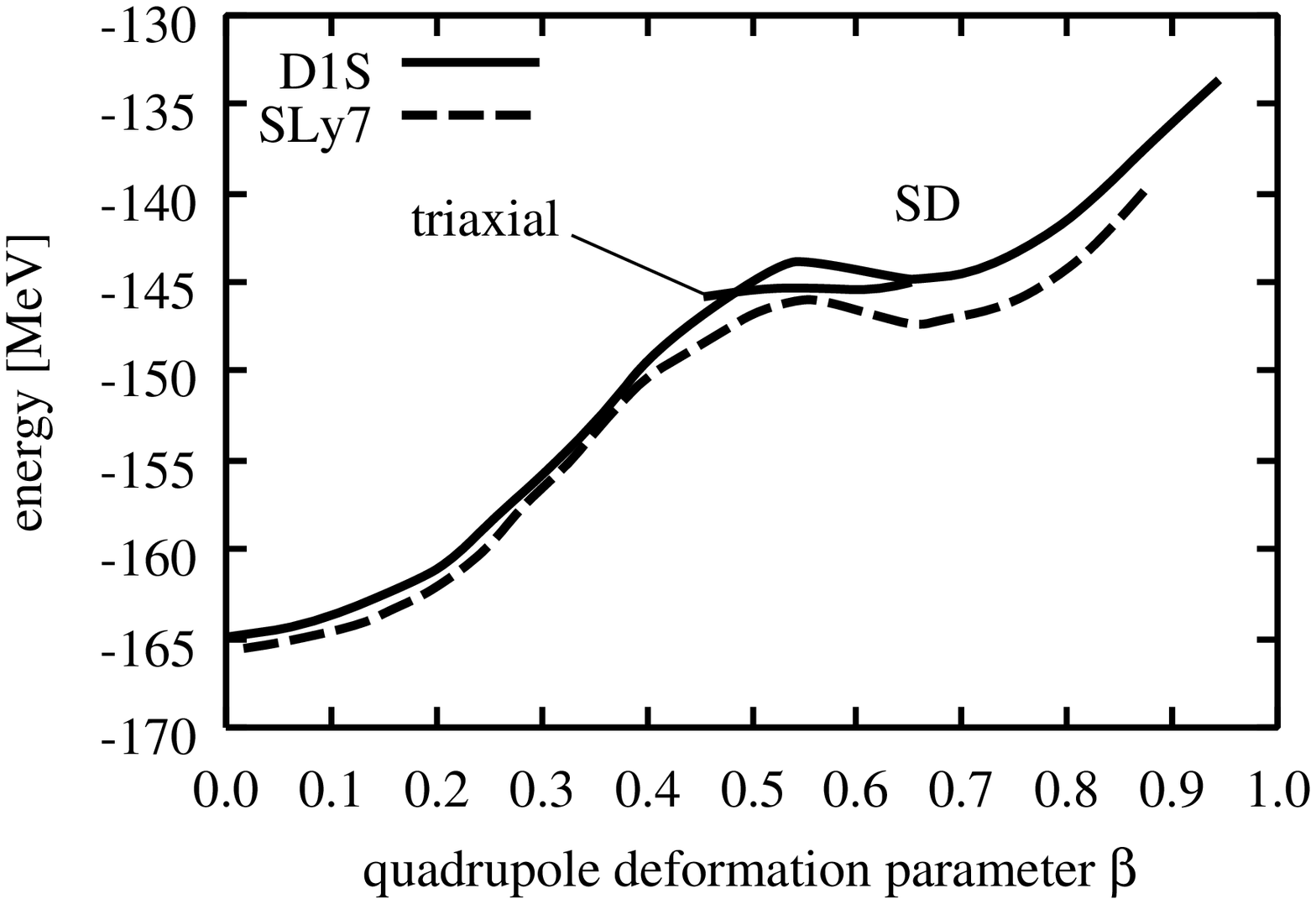} &
      \includegraphics[width=0.3\textwidth]{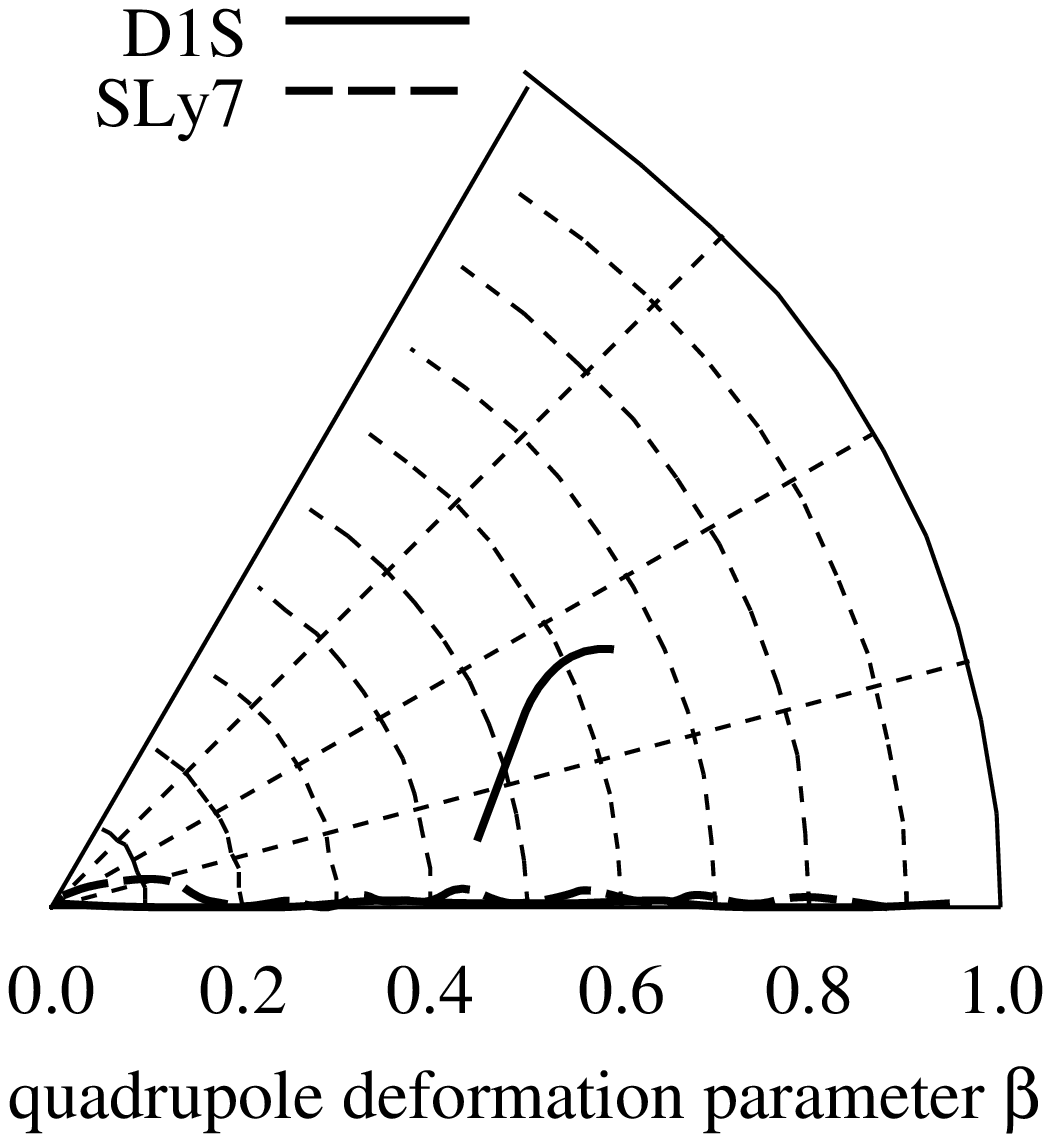}
    \end{tabular}
      \caption{
	(a) Energy curves as functions of quadrupole deformation parameter $\beta$ for positive-parity state in $^{24}$O using D1S and  SLy7 forces. 
	(b) Projection of the energy curves onto $\beta$-$\gamma$ plane. 
      }
      \label{fig:24O_E+}
  \end{center}
\end{figure}

\begin{figure}[tbp]
  \begin{center}
    \begin{tabular}{cc}
      \huge (a) & \huge (b) \\
      \includegraphics[width=0.4\textwidth]{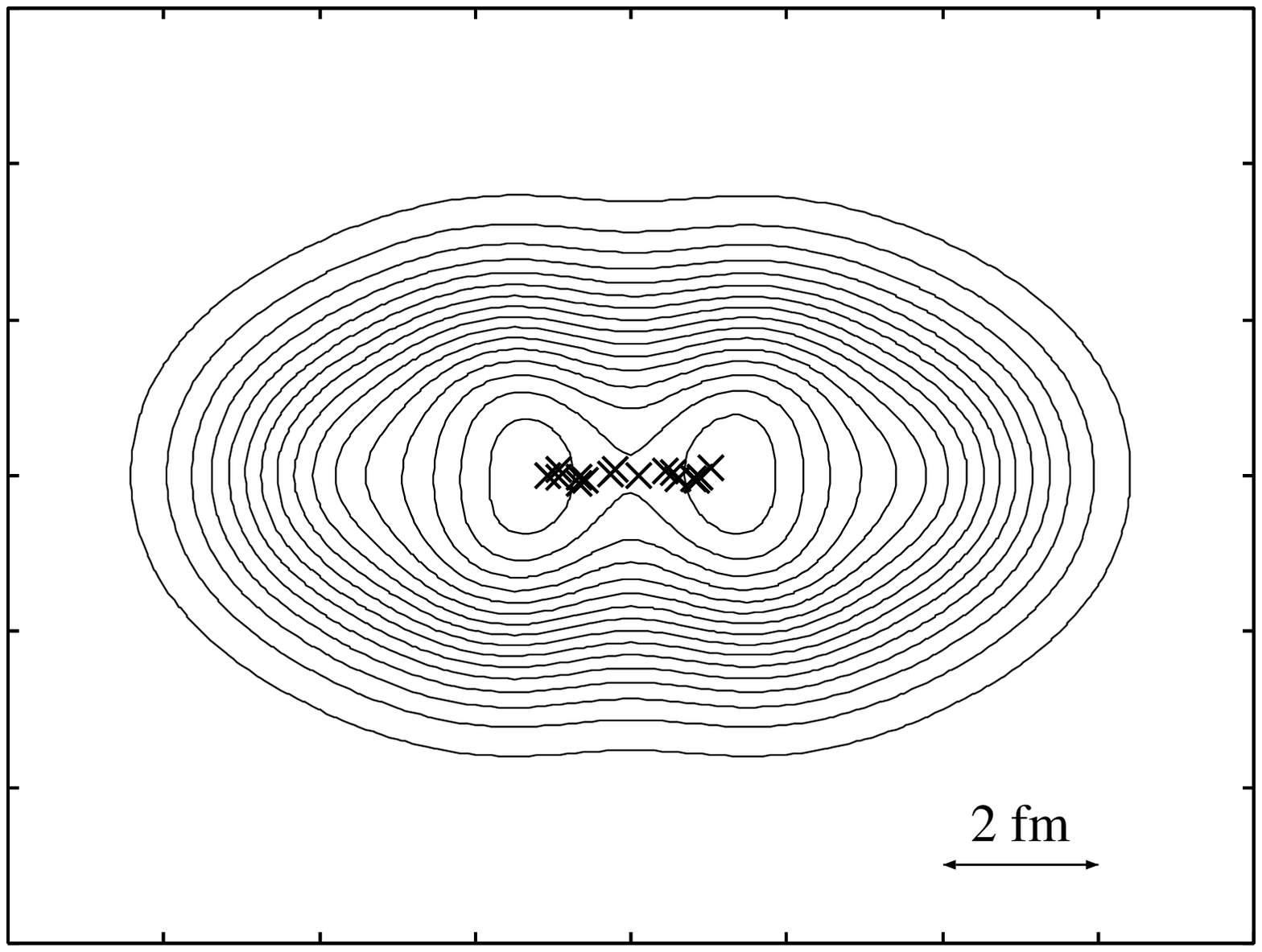} &
      \includegraphics[width=0.4\textwidth]{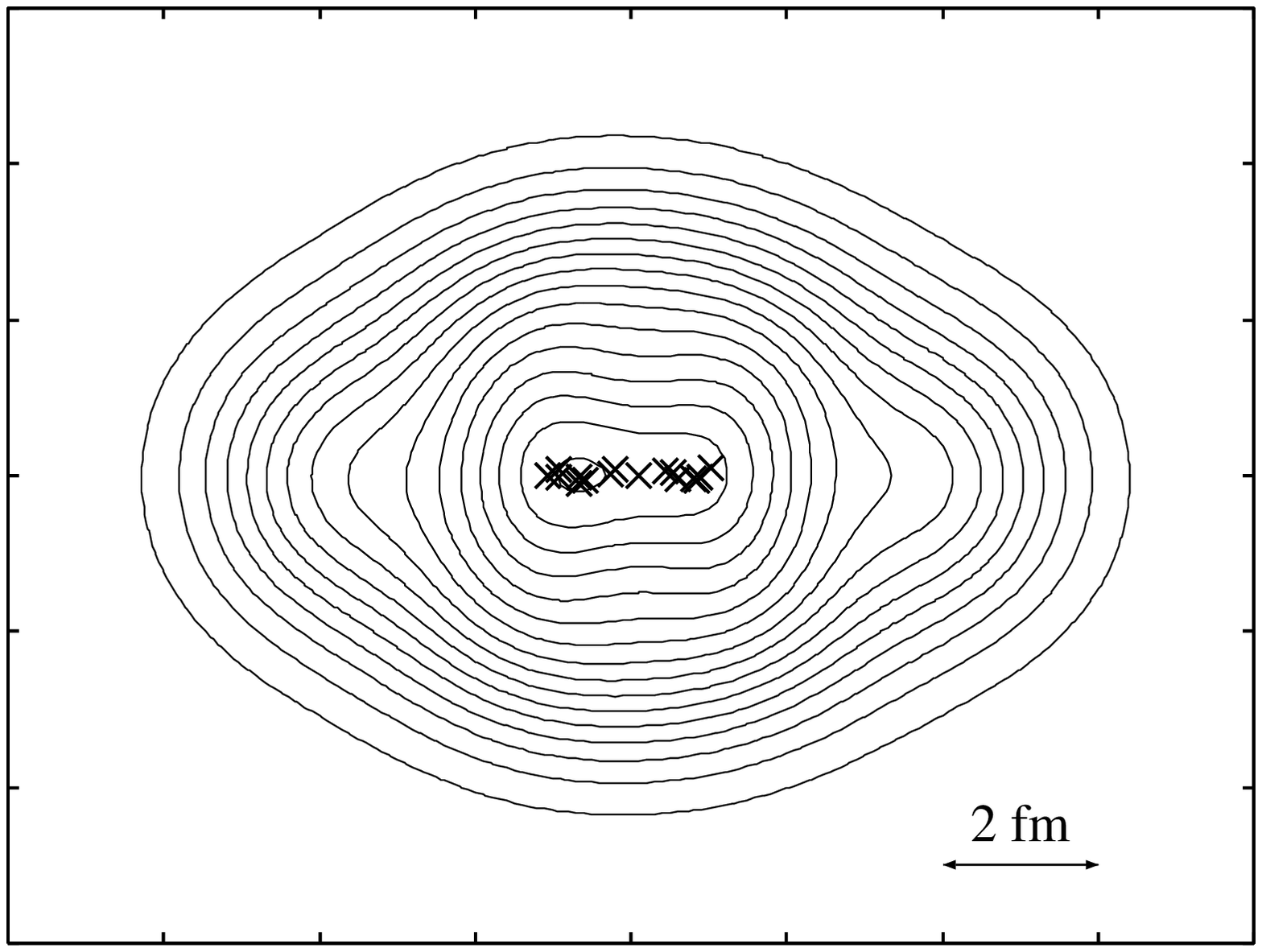}
    \end{tabular}
    \caption{
      (a) Neutron density distribution for an axial symmetric structure at the SD local minimum ($\beta=0.66$) in $^{24}$O. 
      (b) Neutron density distribution for triaxial structure at $\beta=0.65$ in $^{24}$O. 
    }
    \label{fig:24O_density}
  \end{center}
\end{figure}

The energy curves for positive-parity states in neutron-rich nuclei $^{24}$O are shown in Fig.~\ref{fig:24O_E+}(a). 
In the case of the D1S force, prolate and triaxial shapes coexist in the region from $\beta\sim 0.4$ to 0.7, 
whereas the triaxial shape is not obtained in the case of the SLy7 force, as shown in Fig.~\ref{fig:24O_E+}(b). 
The energy curves for the prolate deformations in the cases of the D1S and SLy7 forces are qualitatively similar. 
Those have spherical ground states and local minima at $\beta\sim 0.7$ (the SD local minima). 
The excitation energies of the SD local minima for the D1S and SLy7 forces are 20.0 and 18.1 MeV, respectively. 
Neutron density distributions of the axial symmetric state and the triaxial structure for the D1S force at $\beta\sim 0.65$ are shown in Fig.~\ref{fig:24O_density}(a) and (b), respectively, for the D1S force. 
The neutron density distribution of axial symmetric structure has a neck, and it is similar to that of the SD local minimum in $^{32}$S (Fig.~\ref{fig:32S_SD}), which have the same neutron number $N=16$ as that of $^{24}$O. 

\section{Analysis of clustering aspects}
\label{sec:clustering}
As mentioned in the previous section, it is found that 
the well-developed cluster structure appears 
in the largely deformed states of $p$-shell and $sd$-shell nuclei.
In this section, we discuss clustering correlations 
in the deformed states in more detail.
For this aim, we introduce ideal cluster wave functions
given by the Margenau-Brink-type wave functions and take the
overlap of the deformed-basis AMD wave functions 
with the cluster wave functions.

We also discuss the difference of the results 
between the D1S and the Sly7 forces
from the point of view of the cluster structures.
As discussed before, the D1S and SLy7 forces give quite similar 
energy curves for the mean-field-type structures with no cluster 
structure. This is the case in the 
``MF'' states in $^{24}$Mg and $^{36}$Ar, and those for the
negative-parity states in $^{20}$Ne
(see Sec.~\ref{sec:sd-shell}).
In contrast to those results, the difference between 
the D1S and SLy7 forces is found in the energies of the 
deformed states with the cluster structures. 
The D1S force gives
lower energy for most of the cluster-structure states 
than those of the SLy7 force.
When the cluster structures develop, 
the energy difference tends to become larger. 
Then, it is natural to expect that the interaction 
dependence of the $\beta$ energy curve in the D1S and SLy7 results originates in the 
clustering effects.
Comparing the energies of the ideal cluster wave functions for 
the D1S force with those for the SLy7, 
we will discuss the interaction dependence 
in relation with clustering.
 
In order to discuss the clustering effects in the deformed states, 
we calculate $2\alpha$, equilateral-triangular $3\alpha$, $\alpha$-$^{16}$O and $^{16}$O-$^{16}$O cluster structure components $S_{2\alpha},\  S_{3\alpha},\  S_{\alpha\mbox{-}^{16}{\rm O}}$ and $S_{{\rm ^{16}O\mbox{-}^{16}O}}$, respectively, in the 
deformed-basis AMD wave functions for 
$^8$Be, $^{12}$C, $^{20}$Ne and $^{32}$S, respectively, 
as functions 
of quadrupole deformation parameter $\beta$
using Margenau-Brink-type wave functions, which are defined in Appendix.
The cluster components $S_X$, where $X = 2\alpha,\  3\alpha,\  \alpha\mbox{-}^{16}{\rm O}$ and $^{16}{\rm O}\mbox{-}^{16}{\rm O}$, take $0\leq S_X \leq 1$, and, when $S_X=1$, it means the deformed-basis wave function is described by a traditional cluster model wave function. 
We also calculate the energies of those 
cluster wave functions as functions of the intercluster distance
using the D1S and SLy7 forces.

\begin{figure}[tbp]
  \begin{minipage}{0.5\textwidth}
    \begin{center}
      \includegraphics[width=0.9\textwidth]{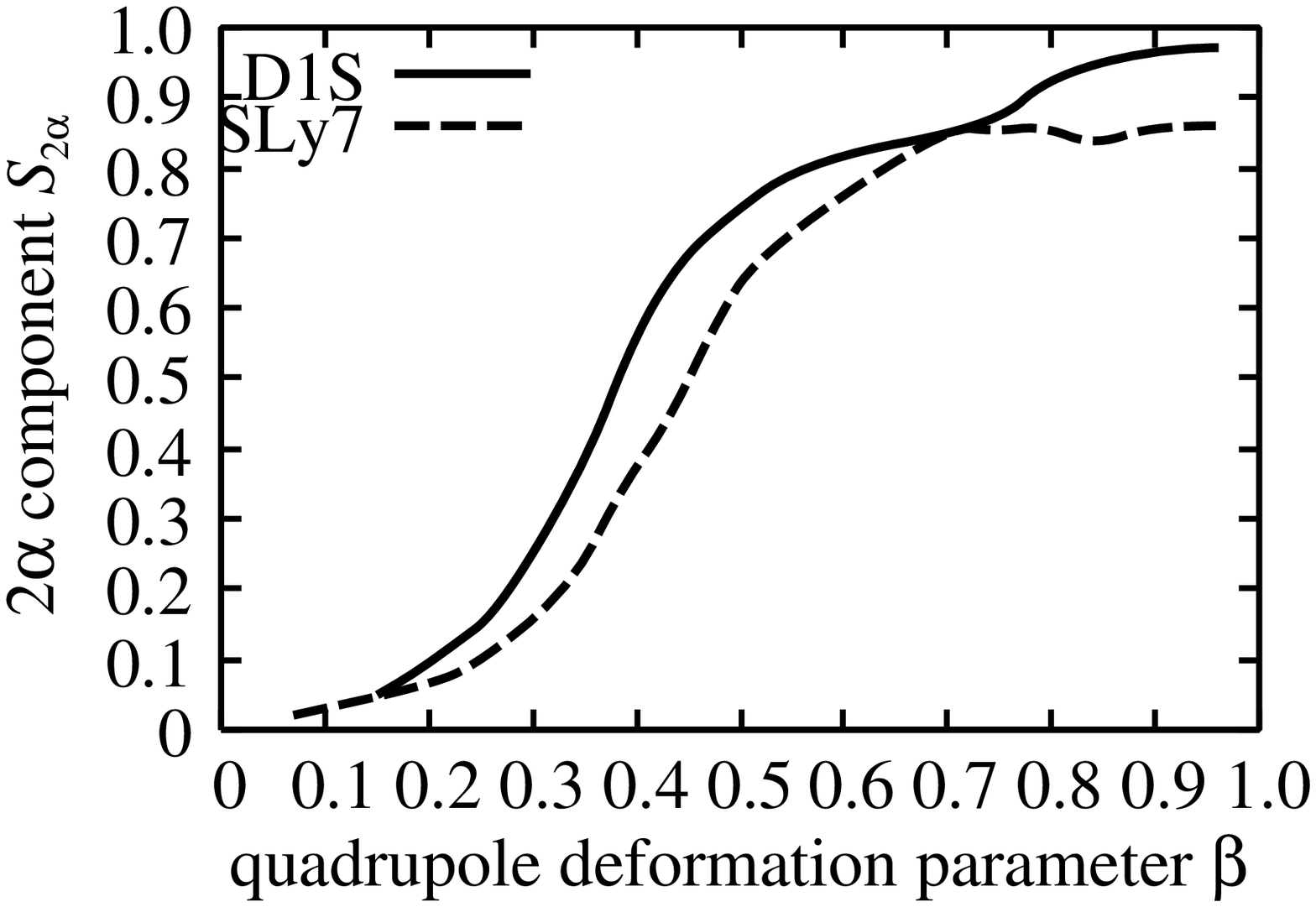}
      \caption{$2\alpha$ components as functions of quadrupole deformation parameter $\beta$ in the cases of the D1S and SLy7 forces. 
	The width parameter $\nu$ is fixed to be $\nu=0.20$ fm$^{-2}$, 
	which is the optimized value for an $\alpha$ cluster 
	with the $(0s)^4$ structure. 
      }
      \label{fig:8Be_2alpha}
    \end{center}
  \end{minipage}
  \begin{minipage}{0.5\textwidth}
    \begin{center}
      \includegraphics[width=0.9\textwidth]{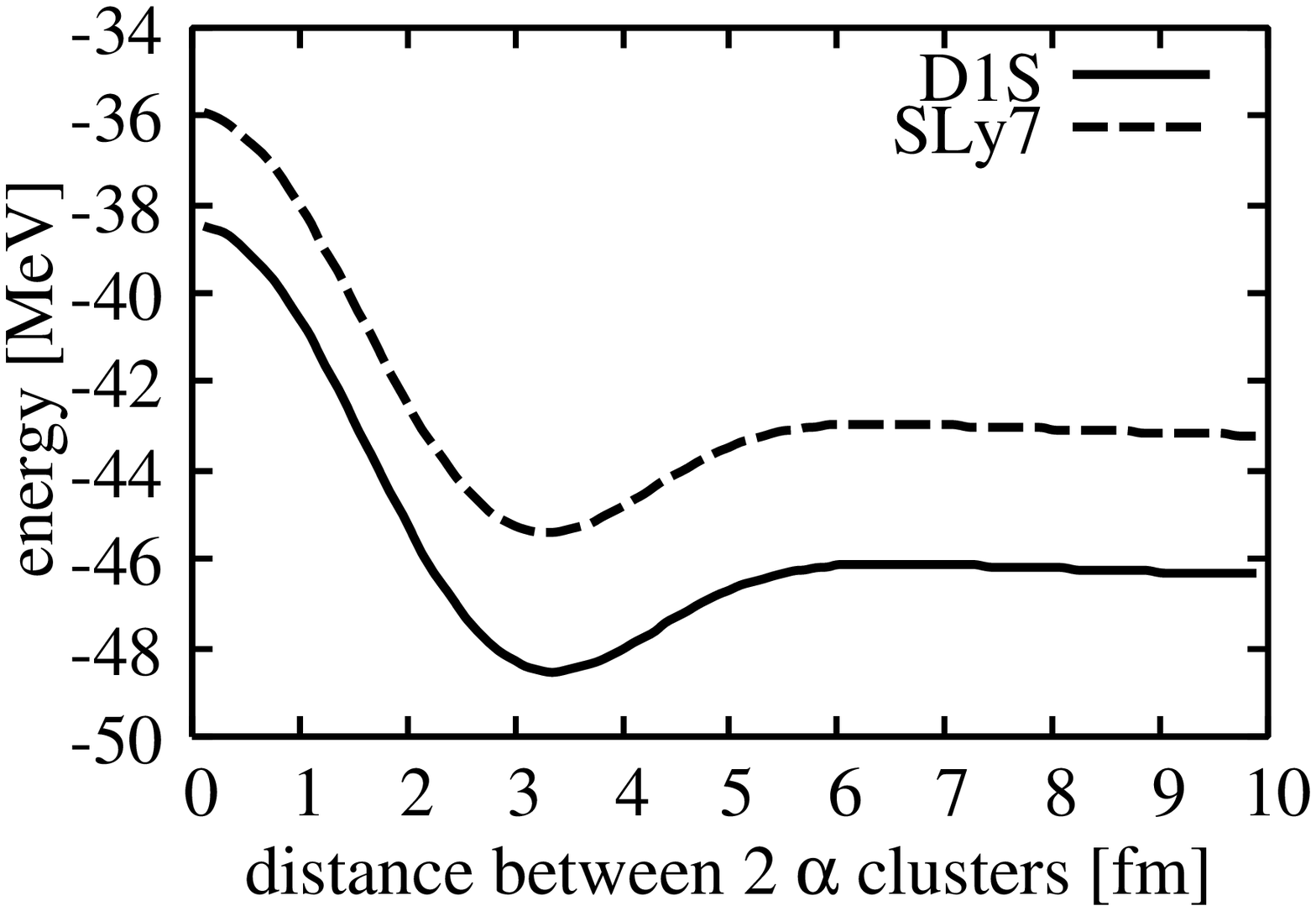} 
      \caption{
	Energy curves of $\alpha$-$\alpha$ Margenau-Brink wave functions for positive-parity states for the D1S and SLy7 forces. 
	Width parameters $\boldsymbol\nu$ are fixed to $\nu_x = \nu_y = \nu_z = 0.20$ fm$^{-2}$. 
      }
      \label{fig:2alpha}
    \end{center}
  \end{minipage}
\end{figure}

Figure~\ref{fig:8Be_2alpha} shows $2\alpha$ cluster structure components $S_{2\alpha}$ in the deformed states of $^8$Be. 
In the deformed states of $^8$Be, the $2\alpha$ cluster 
components are largely contained.
The small $S_{2\alpha}$ value in the small $\beta$ region
indicates the breaking of the $2\alpha$ cluster.
With the increase of the deformation $\beta$, 
the value of $S_{2\alpha}$ increases rapidly 
in $\beta\sim 0.3$ region and shows the cluster formation.
At the minimum ($\beta\sim0.7$) of the energy curve, 
$S_{2\alpha}$ is 0.8 which indicates the $2\alpha$ 
cluster structure of the ground state of $^8$Be. 
Let us show the energies of the $2\alpha$ cluster wave functions
as functions of the distance $d$ between the centers of masses of 
two $\alpha$'s in Fig.~\ref{fig:2alpha}.
The width parameter of the $\alpha$ clusters 
is fixed to be $\nu=0.20$ fm$^{-2}$, 
which is the optimized value for an $\alpha$ nucleus. 
The binding energies of $\alpha$ are 29.7 and 28.1 MeV 
for the D1S and the SLy7 forces, respectively. 
The local minimum is located at almost the same intercluster 
distance $d\sim 3$ fm in both cases of the D1S and SLy7 forces,
but the energies in the case of the D1S force are lower 
than those in the case of the SLy7 force in all  the region of the
intercluster distance. It means that the $2\alpha$ 
cluster structure is
energetically favored more in the case of the D1S force than the
SLy7 force.
Then the difference of the $\beta$ energy curve 
[Fig.~\ref{fig:p-shell}(a)] between the D1S and
the SLy7 forces is interpreted as follows.
With the increase of the deformation $\beta$, 
$2\alpha$ cluster structure component is enhanced.
Once the cluster structure develops, the energies of the 
deformed states are different between the D1S and the SLy7 forces
reflecting the energy difference of the $2\alpha$ cluster
wave function between two forces. As a result, 
the energy curve is steeper in the D1S force,
and the deformation with the developed cluster structure is more 
enhanced in the D1S force than in the SLy7 force.

\begin{figure}[tbp]
  \begin{minipage}{0.5\textwidth}
    \begin{center}
      \includegraphics[width=0.9\textwidth]{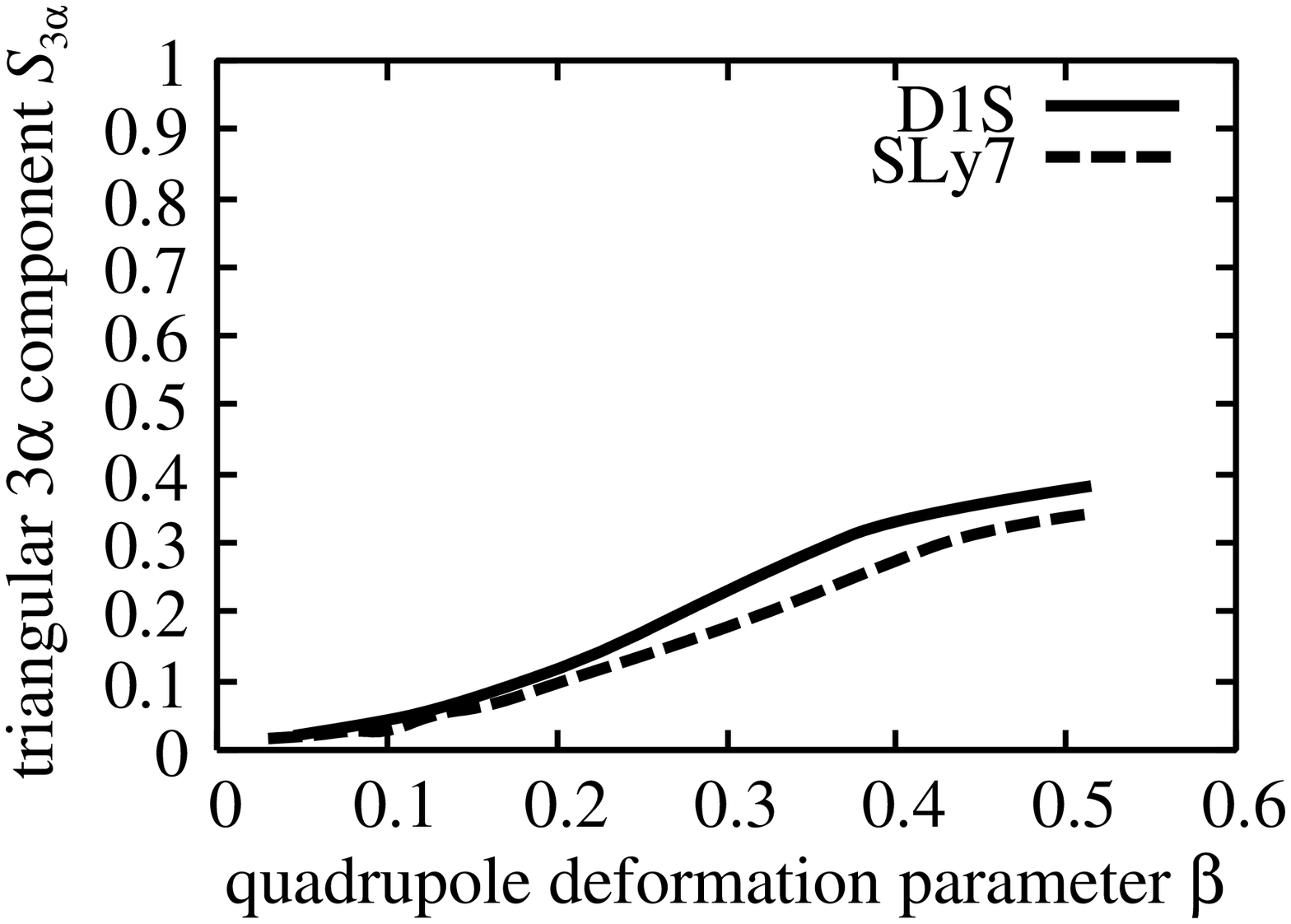}
      \caption{
	Triangular $3\alpha$ cluster structure components in deformed structures of $^{12}$C obtained by energy variations imposing constraint of quadrupole deformation parameter $\beta$. 
	The width parameter $\nu$ is fixed to be $\nu=0.20$ fm$^{-2}$, 
	which is optimized for an $\alpha$ cluster with the $(0s)^4$ structure. 
      }
      \label{fig:12C_3alpha}
    \end{center}
  \end{minipage}
  \begin{minipage}{0.5\textwidth}
    \begin{center}
      \includegraphics[width=0.9\textwidth]{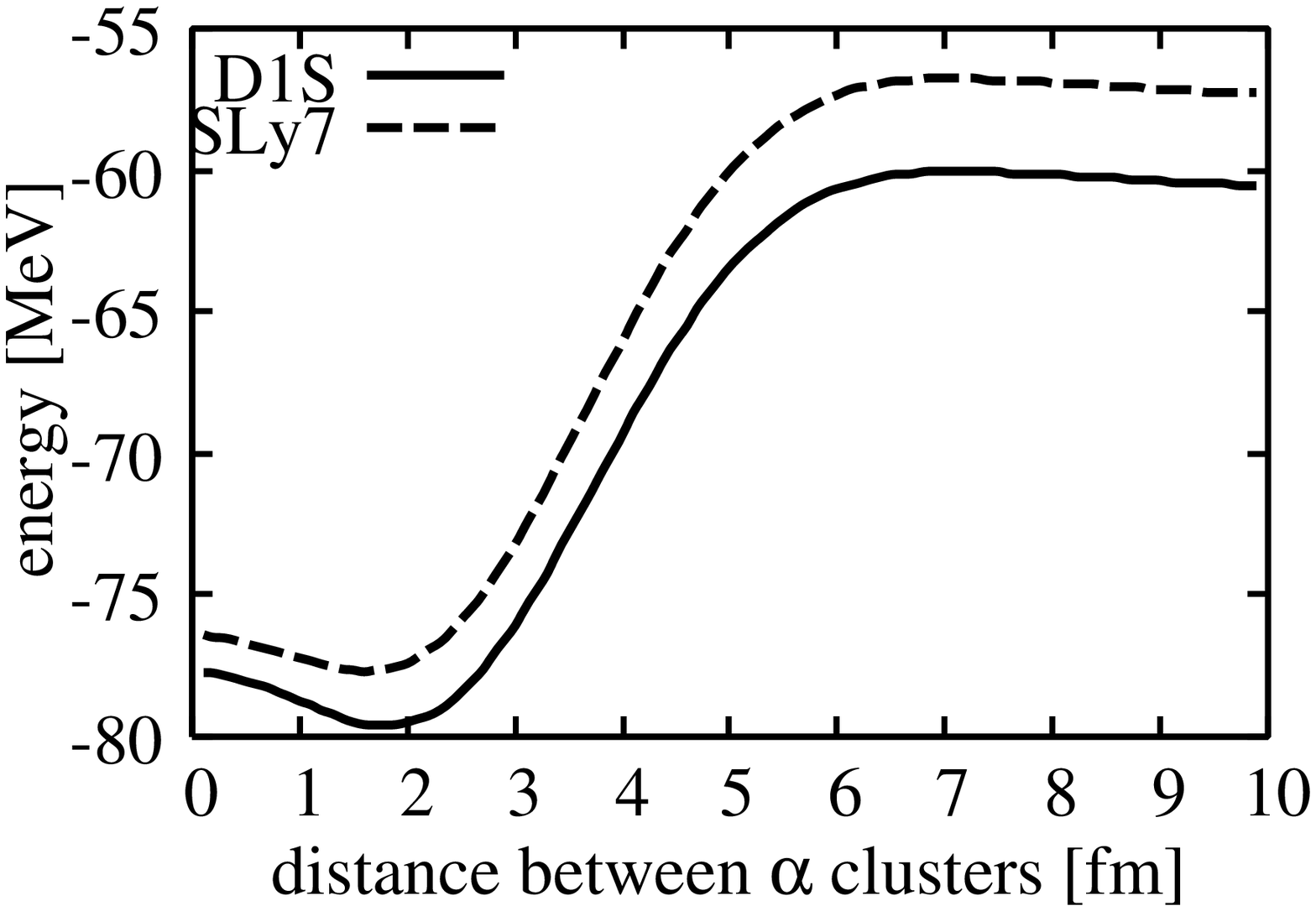} 
      \caption{Energy curve of triangular 3$\alpha$ Margenau-Brink wave functions for positive parity states in the cases of the D1S and SLy7 forces. 
	The width parameter $\nu$ is fixed to be $\nu=0.20$ fm$^{-2}$, 
	which is optimized for an $\alpha$ cluster with the $(0s)^4$ structure. 
      }
      \label{fig:3alpha}
    \end{center}
  \end{minipage}
\end{figure}

In the deformed states of $^{12}$C, triangular $3\alpha$ cluster structure components $S_{3\alpha}$ are contained as shown in Fig.~\ref{fig:12C_3alpha}. Here the cluster structure component is calculated by
using the Margenau-Brink-type wave functions 
with the equilateral-triangle configurations.
In the spherical region, the $3\alpha$ cluster structure component 
is almost zero, and the component becomes larger as increasing the values of $\beta$. The $S_{3\alpha}$ is approximately 0.4 in the
$\beta\sim 0.5$ region in both cases of the D1S and SLy7 forces. 
This means that the oblately deformed states of $^{12}$C still 
contain non-cluster structure components as well as the $3\alpha$ cluster 
components.
In Fig.~\ref{fig:3alpha}, the energy of the $3\alpha$ cluster 
wave functions projected to the positive party states 
as a function of the intercluster distance.
Similarly to the case of the $2\alpha$ cluster wave functions, 
the energy for the D1S force is lower than that for the SLy7 force.
We consider that this energy difference is one of the reasons 
why the D1S and SLy7 forces give the different shapes of the ground
state: the former force gives the oblate ground state 
and the latter shows the spherical one.
That is to say that 
the oblate deformation is favored in the case of the D1S force
because it has the larger energy gain due to 
the $3\alpha$ clustering.

\begin{figure}[tbp]
  \begin{center}
  \begin{tabular}{cc}
    {\huge (a) positive} & {\huge (b) negative}\\
    \includegraphics[width=0.45\textwidth]{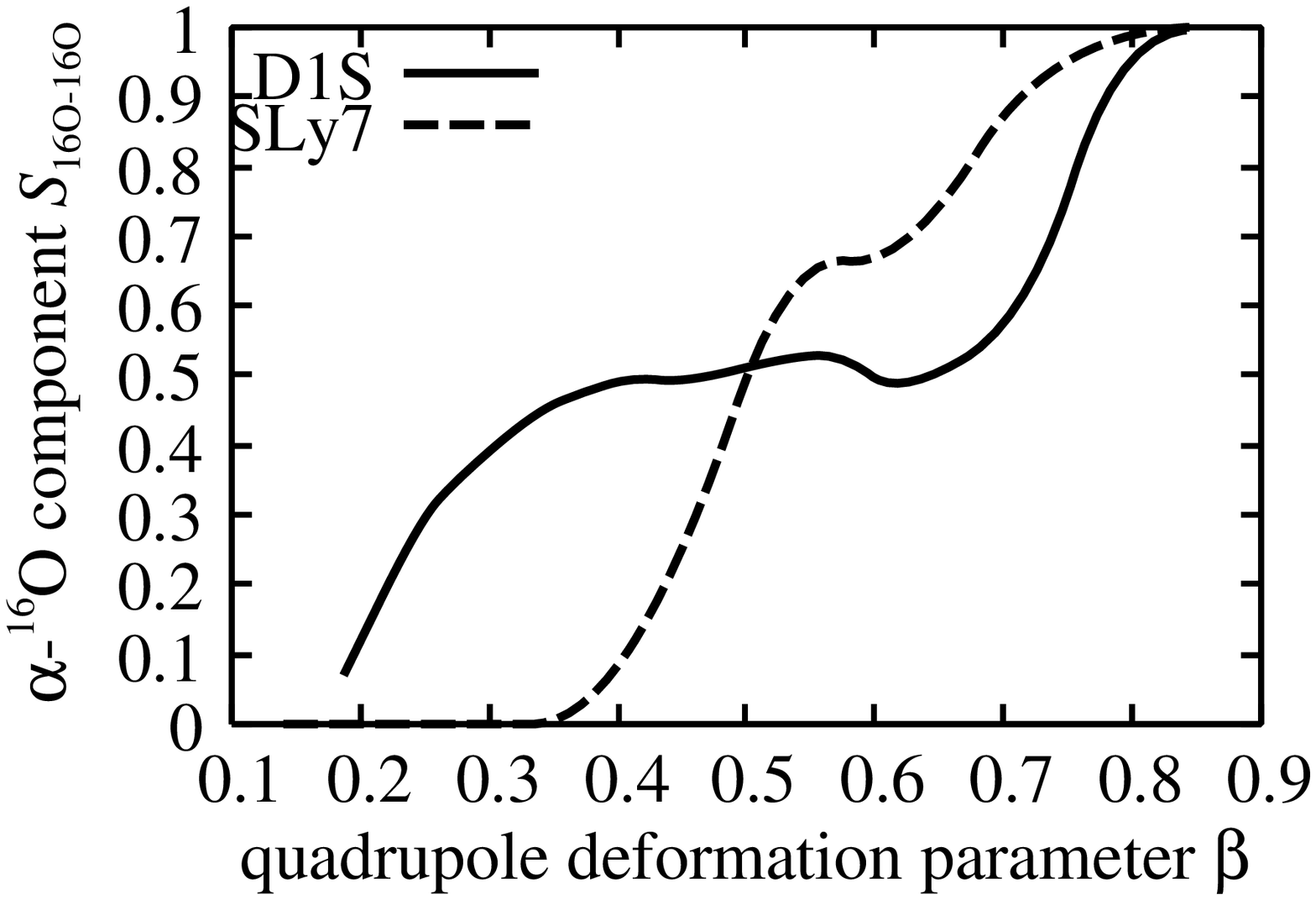} &
    \includegraphics[width=0.45\textwidth]{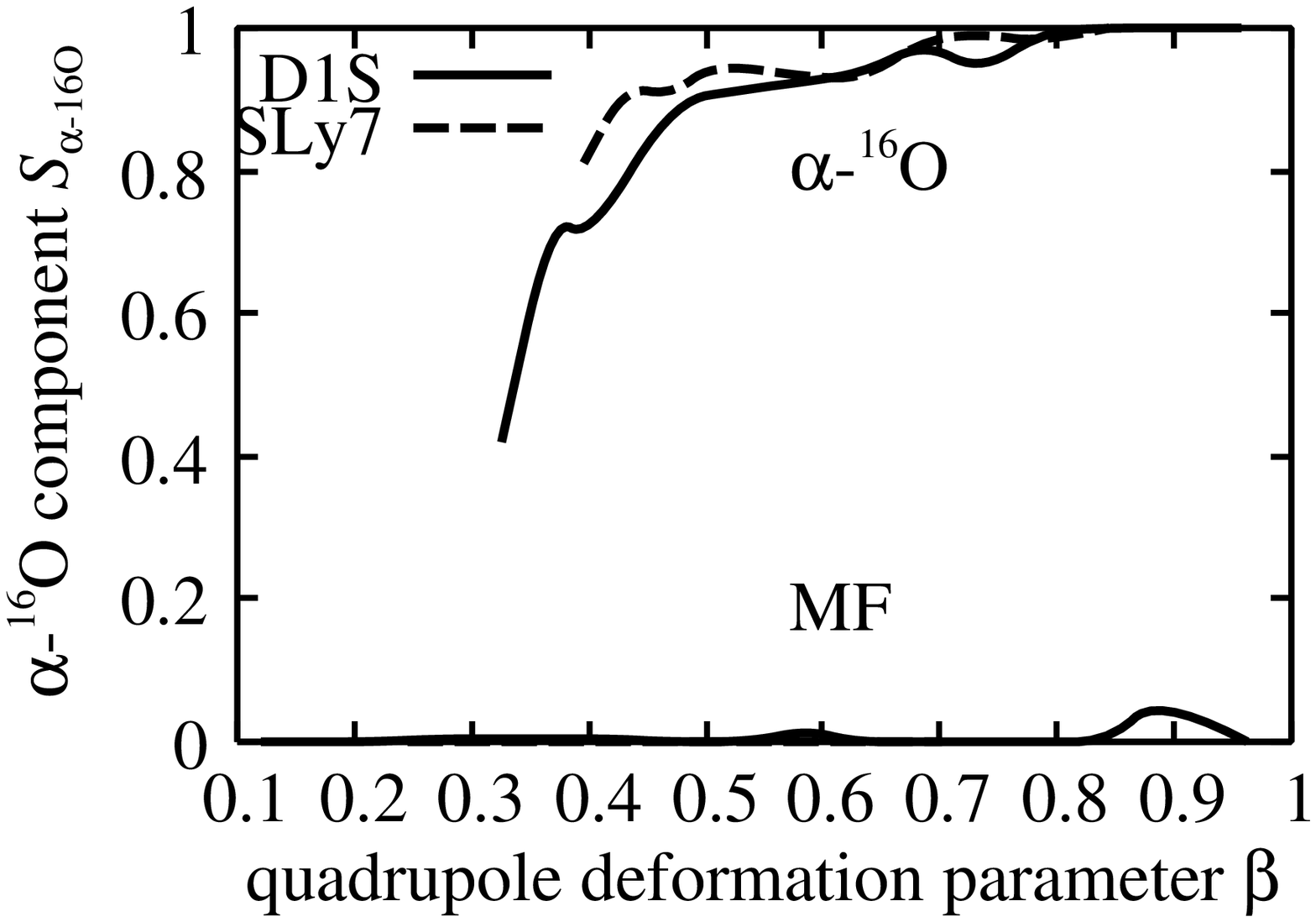}
  \end{tabular}
  \caption{
    $\alpha$-$^{16}$O cluster structure components in deformed structures of $^{20}$Ne obtained by energy variations imposing constraint of quadrupole deformation parameter $\beta$. 
    In the cases of (a) positive and (b) negative parity states are presented. 
    Width parameters $\boldsymbol\nu$ are fixed to $\nu_x=\nu_y=\nu_z=0.16$ fm$^{-2}$. 
  }
  \label{fig:S_4He16O}
  \end{center}
\end{figure}

\begin{figure}[tbp]
  \begin{center}
    \begin{tabular}{cc}
      \huge (a) positive & \huge (b) negative\\
      \includegraphics[width=0.45\textwidth]{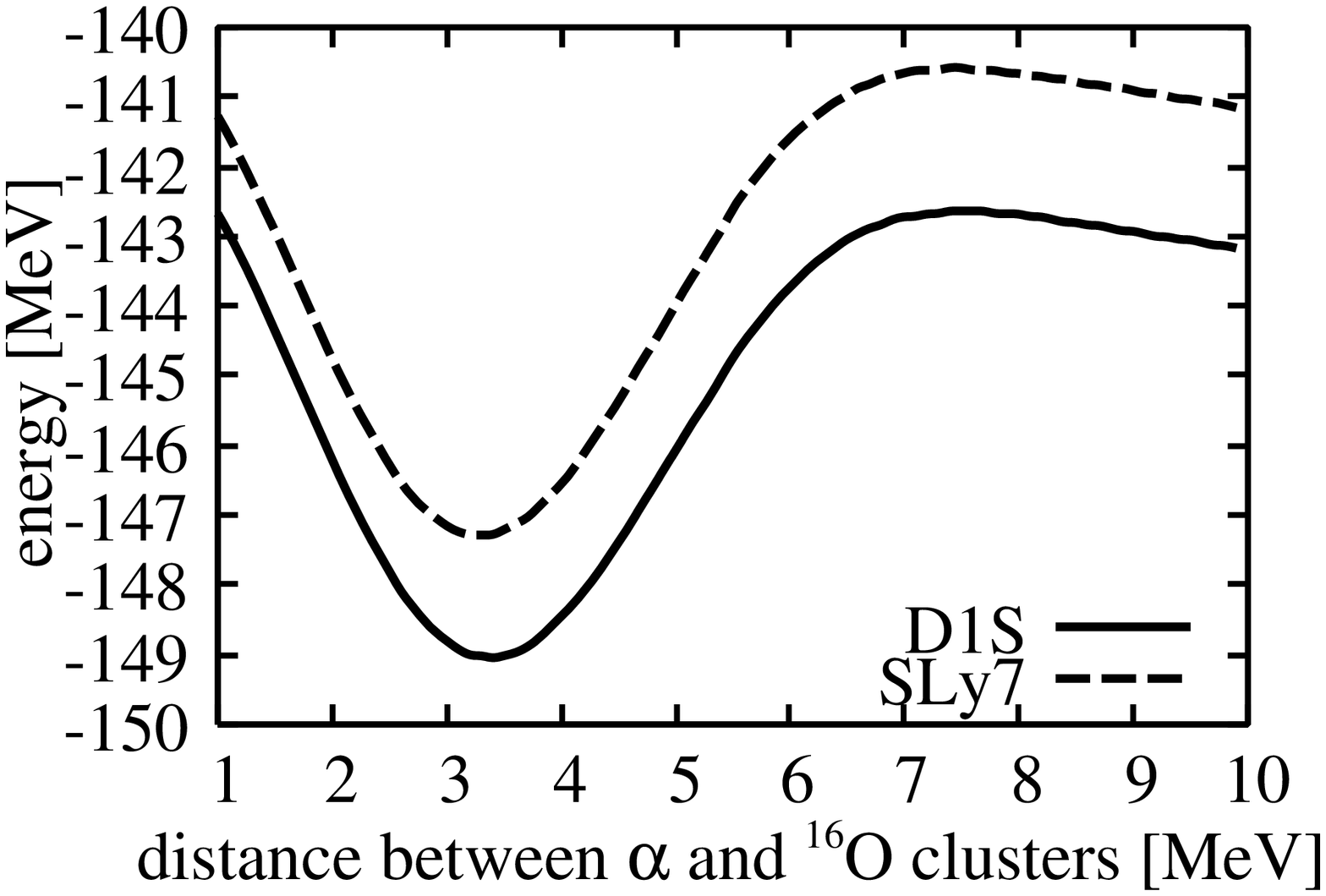} &
      \includegraphics[width=0.45\textwidth]{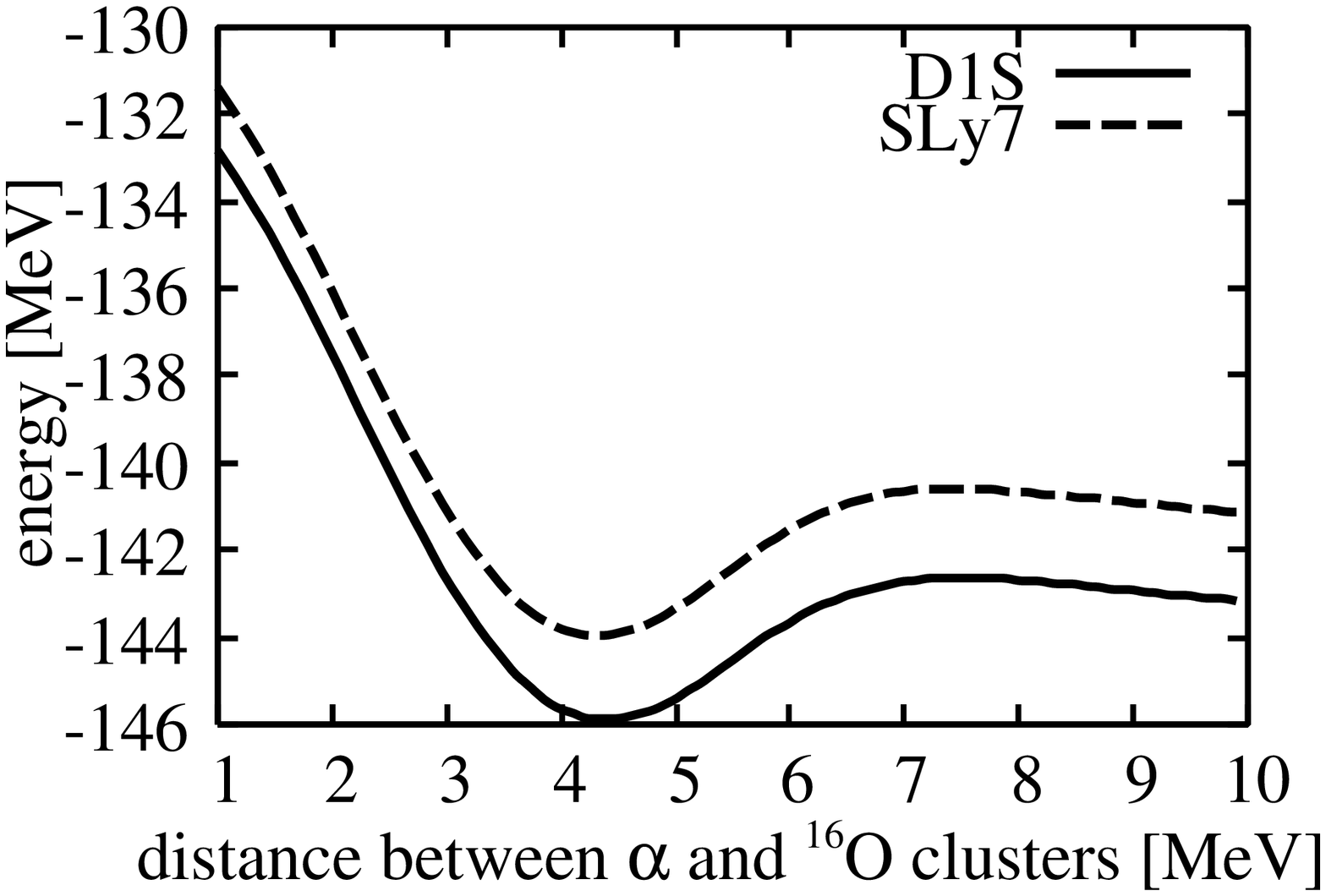}
    \end{tabular}
    \caption{
      Energy curves of $\alpha$-$^{16}$O Margenau-Brink-type cluster wave functions for (a) positive- and (b) negative-parity states in the cases of the D1S and SLy7 forces. 
      Width parameters $\boldsymbol\nu$ are fixed to $\nu_x=\nu_y=\nu_z=0.16$ fm$^{-2}$. 
    }
    \label{fig:a-16O}
  \end{center}
\end{figure}

Figures ~\ref{fig:S_4He16O}(a) and (b) show the $\alpha$-$^{16}$O cluster
component $S_{\alpha\mbox{-}{\rm ^{16}O}}$ 
in the deformed states of $^{20}$Ne 
with positive and negative parity, respectively. 
The width parameter of the $\alpha$ and the $^{16}$O clusters
are fixed to $\nu=0.16$ fm$^{-2}$, 
which is the optimized value for $^{20}$Ne in the case of spherical wave packets. 
As shown in Fig.~\ref{fig:sd-shell}(a), 
the energy minimum state for the ground state 
is the deformed state with positive parity 
at $\beta\sim 0.4$. 
In the case of the D1S force,
the $\alpha$-$^{16}$O cluster structure component is contained 
50 \% in the energy minimum state. On the other hand, in the
case of the SLy7 force 
the $\alpha$-$^{16}$O cluster structure component is strongly 
suppressed in the $\beta\le 0.4$ region. The SLy7 force shows
the rapid change of $S_{\alpha\mbox{-}{\rm ^{16}O}}$ around $\beta\sim 0.5$, which 
indicates the drastic transition from the mean-field-type deformed
structure into the cluster structure.
The significant $\alpha$-$^{16}$O cluster structure component of the ground
state obtained with the D1S force is consistent with the 
fact that the ground state of $^{20}$Ne have 
strong $\alpha$-$^{16}$O correlations.\cite{hor72,fuj80,kim04a}. 
The larger cluster structure component in the case of the D1S force
than the SLy7 force is understood by the lower energy 
of the $\alpha$-$^{16}$O cluster wave function in the
D1S force than the SLy7 [Fig.~\ref{fig:a-16O}(a)].

In the deformed states with negative parity of $^{20}$Ne,
two types of structures are obtained: the mean-field-type structure
and the $\alpha$-$^{16}$O cluster structure which are
labeled ``MF'' and ``$\alpha$-$^{16}$O'', respectively, as mentioned before.
As seen in Fig.~\ref{fig:S_4He16O}(b), the $\alpha$-$^{16}$O-like
deformed states are dominated by the $\alpha$-$^{16}$O cluster
components. In contrast, the MF states have no $\alpha$-$^{16}$O
cluster structure component even in the largely deformed states.

Figures~\ref{fig:a-16O}(a) and (b) show the energies of the
$\alpha$-$^{16}$O cluster wave functions
projected to the positive- and negative-parity states, respectively, as functions of the distance between $\alpha$ 
and $^{16}$O clusters in the cases of the  D1S and SLy7 forces. 
In both the positive- and negative-parity states, 
the energies for the D1S force are lower 
than those for the SLy7 force as in the cases of $2\alpha$ 
and $3\alpha$ cluster wave functions.
Reflecting this energy difference 
of the cluster wave function between
the D1S and the SLy7 forces, 
the energies of the deformed states with clustering 
obtained by the $\beta$-constraint are relatively 
lower in the case of the D1S than the case of the SLy7 force.
In particular, those for the negative-parity states 
are almost consistent with the energies of the negative-parity 
cluster wave functions.
This results in the lower excitation energy of the 
negative-parity states with the cluster correlation
which constructs the $K^\pi=0^-$ band
in the $^{20}$Ne.

\begin{figure}[tbp]
  \begin{minipage}{0.5\textwidth}
    \begin{center}
      \includegraphics[width=0.9\textwidth]{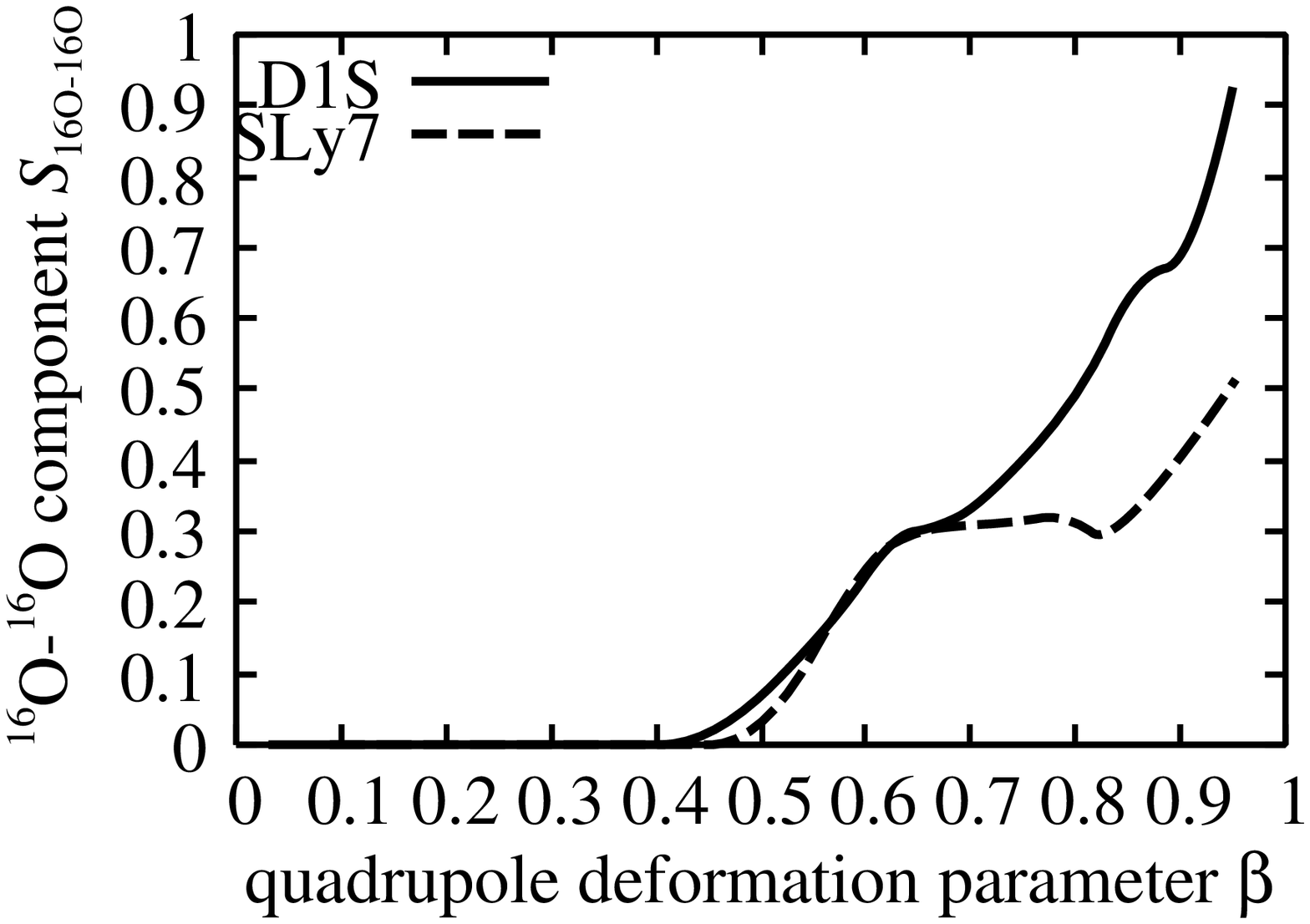}
      \caption{
	$^{16}$O-$^{16}$O cluster structure components in deformed structures of $^{32}$S obtained by energy variations imposing constraint of quadrupole deformation parameter $\beta$. 
      }
      \label{fig:S_16O16O}
    \end{center}
  \end{minipage}
  \begin{minipage}{0.5\textwidth}
    \begin{center}
      \includegraphics[width=0.9\textwidth]{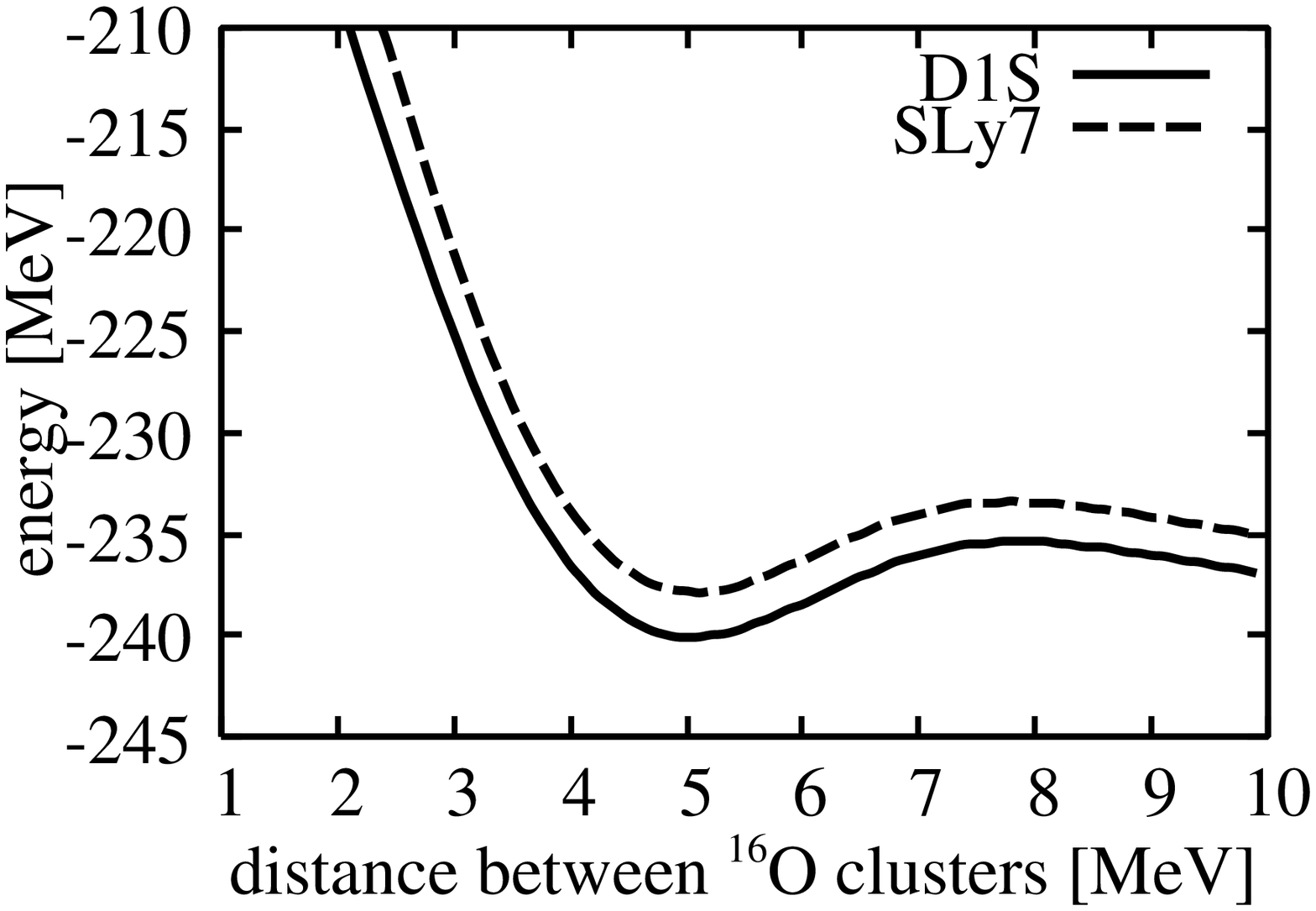} 
      \caption{Energy curves of $^{16}$O-$^{16}$O Margenau-Brink wave functions for positive parity states in the cases of the D1S and SLy7 forces. 
	Width parameters $\boldsymbol\nu$ are fixed to $\nu_x=\nu_y=\nu_z=0.16$ fm$^{-2}$. }
      \label{fig:16O-16O}
    \end{center}
  \end{minipage}
\end{figure}

Figure~\ref{fig:S_16O16O} shows $^{16}$O-$^{16}$O cluster structure components 
$S_{{\rm ^{16}O\mbox{-}^{16}O}}$ in the deformed states of $^{32}$S.
The width parameter of the $^{16}$O cluster is 
fixed to $\nu=0.16$ fm$^{-2}$, 
which is the optimized value for a $^{16}$O cluster.
In the small deformation region ($\beta\le 0.4$), 
the $^{16}$O-$^{16}$O cluster structure component 
is almost zero, while it is significantly contained in 
the largely deformed states($\beta\ge 0.6$).
The value of $S_{{\rm ^{16}O\mbox{-}^{16}O}}$ of
the local minimum for the SD state 
at $\beta\sim 0.7$ is 0.3, which indicates the
$^{16}$O-$^{16}$O cluster correlation in the SD state\cite{kim04b}. 
Figure~\ref{fig:16O-16O} shows the energy curves projected to positive party as a function of the distance between two $^{16}$O clusters for the D1S and SLy7 forces. The D1S force gives
the lower energy of the cluster wave function
than the SLy7 forces. 
It is consistent that the energy of the SD local minimum on the $\beta$ energy curve
for the D1S force is slightly lower than that for the
SLy7 force, and the SD state in case of D1S force contains larger $^{16}$O-$^{16}$O cluster structure component $S_{{\rm ^{16}O\mbox{-}^{16}O}}$.

The present results show that 
the cluster structures affect the deformed states 
in $sd$-shell nuclei as well as $p$-shell nuclei. 
Although the D1S and SLy7 forces give the similar energy curves 
for the mean-field-type structures, 
they give the different energies 
for the ideal cluster wave functions. In the $p$-shell and 
$sd$-shell regions, the D1S force gives the lower energies 
of the cluster states than those of the SLy7 force. 
This difference is reflected in the excitation energies of the
largely deformed states with clustering correlation.
In other words, the energy of the 
cluster structures are reflected on the $\beta$ 
energy curves due to the clustering effect in the deformed 
states.

\section{Conclusion}
\label{sec:conclusion}
We have studied deformations and clustering correlations in the
$p$-shell and $sd$-shell regions using the Gogny D1S and Skyrme SLy7 forces. 
The effective-interaction dependence has been also discussed. 
We have obtained the energy curves as functions of a quadrupole 
deformation parameter $\beta$ with a parity-projected 
energy variation imposing a constraint of 
quadrupole deformation parameter $\beta$ to the deformed-basis
AMD wave functions. 
We have found that the various deformed structures appear 
with the increase of the deformation $\beta$ 
in both the cases of the D1S and SLy7 forces, which  
give qualitatively similar results. 

We have found that 
various cluster structures appear in $p$- and $sd$-shell 
regions.
For example, in the large $\beta$ region of $N=Z\leq 12$ nuclei, 
the well developed $\alpha$ cluster structures appear 
in both the cases of the D1S and SLy7 forces, 
and the $^{16}$O-$^{16}$O clustering is found 
in the SD local minimum of the $\beta$ energy curves 
of $^{32}$S. 
On the other hand, no cluster structure is seen in the deformed
states in $^{36}$Ar and $^{24}$O. 

In comparison of the results between the D1S and the SLy7 
forces, the properties of the $\beta$ energy curves are qualitatively similar except for $^{24}$O, 
and the $\gamma$ values are also similar 
except for $^{16}$O and $^{24}$O. 
In $^{16}$O, the two forces give different $\gamma$ values in the large $\beta$ region because of the different shapes of $^{12}$C clusters. 
In $^{24}$O, the triaxial deformation appears in the $\beta$ 
region between the spherical shape for the ground state 
and the SD local minimum 
in the case of the D1S force, 
whereas the SLy7 force gives only the axial symmetric shapes.
Cluster structures in the deformed states for the D1S and the SLy7 forces
are also qualitatively similar to each other
in most cases.

For the mean-field-type structures in the $sd$-shell region, 
the two forces give quantitatively quite similar results. 
In contrast, we have found the strong effective interaction 
dependence in $p$-shell and the deformed states with strong
cluster correlations in $sd$-shell nuclei. 
For example, the quantitative difference is found in $^{12}$C, 
for which the D1S force gives the oblate 
shape while the SLy7 force gives the spherical shape 
at the energy minimum in the $\beta$ energy curve.
The cluster structures in the deformed states of
 $^8$Be, $^{12}$C, $^{20}$Ne and $^{32}$S are analyzed 
using $2\alpha$, $3\alpha$, $\alpha$-$^{16}$O and $^{16}$O-$^{16}$O Margenau-Brink-type cluster wave functions, respectively.
Although the D1S and SLy7 forces give the similar energy curves 
for the mean-field-type structures, 
they give the different energies 
for the ideal cluster wave functions. In the $p$-shell and 
$sd$-shell regions, the D1S force gives the lower energies 
of the cluster states than those of the SLy7 force. 
This difference is reflected in the excitation energies of the
largely deformed states with clustering correlation.

In conclusion, various cluster structures 
appear in $sd$-shell region as well as $p$-shell region.
In studying nuclear deformations, cluster structures and 
properties of lighter nuclei should be taken into account, 
because they can affect the nuclear deformations through the
cluster correlation in the deformed states.

\section*{Acknowledgment}
The authors thank to Prof.~Horiuchi for fruitful discussions. 
The numerical calculations has been carried out on SX8 at YITP in Kyoto University and RCNP in Osaka University. 
This work has been supported by a JSPS Research Fellowships for Young Scientists and a Grant-in-Aid for the 21st Century COE ``Center for Diversity and Universality in Physics'' from the Ministry of Education, Culture, Sports, Science and Technology of Japan (MEXT). 

\appendix
\section{Definition of a cluster structure component}
In present paper, cluster structure components $S_X$, where $X = 2\alpha,\ 3\alpha,\ \alpha{\rm \mbox{-}^{16}O}$ or ${\rm ^{16}O\mbox{-}^{16}O}$, are evaluated in order to discuss clustering correlation. 
In this section, we define the cluster structure component $S_X$
in analogy of cluster probability of AMD wave function
in Refs.~\citen{KanadaEn'yo:2003ue,kim04b}. 

The normalized $\pi$-parity projected deformed-basis AMD wave function $\ket{\Phirm^\pi}$ is generally divided into $X$ cluster component $\ket{\Phirm_X^\pi}$ and the residual part $\ket{\Phirm_{R_X}^\pi}$ as
\begin{equation}
  \ket{\Phirm^\pi} = c\ \ket{\Phirm_X^\pi} + \sqrt{1 - |c|^2}\ \ket{\Phirm_{R_X}^\pi},  
\end{equation}
here 
\begin{equation}
  \ovlp{\Phirm_X^\pi}{\Phirm_{R_X}^\pi} = 0. 
\end{equation}
We introduce the projection operator $\hat{P}_X$ which projects out the $X$ cluster component from the $\ket{\Phirm^\pi}$, 
\begin{equation}
  \hat{P}_X \ket{\Phirm^\pi} = c\ \ket{\Phirm_X^\pi} = c \hat{\cal A}\ket{\chi \prod_i \phi(X_i)}. 
\end{equation}
where $\phi(X_i)$ is the normalized internal wave function of the cluster $X_i$ which is contained in $X$ cluster structure and $\chi$ is the wave function of relative motion of the clusters. 

The projection operator $\hat{P}_X$ is defined as following. 
Suppose the wave functions $\ket{\Phirm_i^X}$ that span the functional space of $X$ cluster structure. 
The wave function $\ket{\Phirm_i^X}$ is a Margenau-Brink-type wave function with intercluster distance $d_i$, which is 
\begin{eqnarray}
  &&d_i = d_{\mbox{min}} + i \Delta d\ \ \ \ (0 \leq i \leq i_{\mbox{max}}), \\
  &&\Delta d = \frac{d_{\mbox{max}} - d_{\mbox{min}}}{i_{\mbox{max}}}, 
\end{eqnarray}
where $d_{\mbox{min}}$ and $d_{\mbox{max}}$ are minimum and maximum values of intercluster distance, respectively. 
In present study, $d_{{\rm min}}$ and $\Delta d$ are adopted to $d_{{\rm min}} \sim 0.1$ fm and $\Delta d \sim 1$ fm, respectively. 
Orthogonalized wave function $\ket{\tilde{\Phirm}_\alpha^{X\pi}}$ is obtained by performing unitary transformation from $\ket{\Phirm_i^{X \pi}}$ which is $\pi$-parity projected state of $\ket{\Phirm_i^X}$, 
\begin{eqnarray}
  &&\ket{\tilde{\Phirm}_\alpha^{X \pi}} = u_{\alpha i} \ket{\Phirm_i^{X\pi}}, \\
  && \ovlp{\tilde{\Phirm}_\alpha^{X \pi}}{\tilde{\Phirm}_\beta^{X \pi}} = \delta_{\alpha \beta}. 
\end{eqnarray}
Using the set of orthonormalized wave functions $\ket{\tilde{\Phirm}_\alpha^{X \pi}}$, the projection operator $\hat{P}_X$ to project out the $X$ cluster cluster component is defined as  
  \begin{equation}
    \hat{P}_X = \sum_\alpha \ket{\tilde{\Phirm}_\alpha^{X \pi}}\bra{\tilde{\Phirm}_\alpha^{X \pi}}. 
  \end{equation}
  
Using the projection operator $\hat{P}_X$, cluster components $S_X$ is defined as 
\begin{equation}
  S_X \equiv |c|^2 = \bra{\Phirm^\pi} \hat{P}_X \ket{\Phirm^\pi} = \sum_\alpha |\ovlp{\tilde{\Phirm}_\alpha^{X \pi}}{\Phirm^\pi}|^2. 
\end{equation}
where wave functions $\ket{\tilde{\Phirm}_\alpha^{X}}$ and $\ket{\Phirm}$ are rotated so as to diagonalize moment of inertia matrices as 
\begin{equation}
  \langle x^2 \rangle \leq \langle y^2 \rangle \leq \langle z^2 \rangle. 
\end{equation}
Since $\ket{\tilde{\Phirm}_\alpha^{X \pi}}$ are orthonormalized, $S_X$ takes $0\leq S_X \leq 1$. 


\end{document}